\documentclass[twocolumn,aps,prl,superscriptaddress,notitlepage,longbibliography]{revtex4}
\usepackage[colorlinks=true,urlcolor=blue,citecolor=blue,linkcolor=blue]{hyperref}
\usepackage{graphicx}  
\usepackage{dcolumn}   
\usepackage{bm}        
\usepackage{CJKutf8}
\usepackage{amssymb}
\usepackage{amsfonts}
\usepackage{doi}
\usepackage{verbatim}
\usepackage{epstopdf}
\usepackage{lineno}
\usepackage{amsmath}
\usepackage{braket}
\usepackage{hyperref}

\begin{document}
\title{Dissipative Chiral Channels, Ohmic Scaling and Half-integer Hall Conductivity from the Relativistic Quantum Hall Effect}
\author{Humian Zhou}
\affiliation{International Center for Quantum Materials, School of Physics, Peking University, Beijing 100871, China}
\author{Chui-Zhen Chen}
\email{czchen@suda.edu.cn}
\affiliation{School of Physical Science and Technology, Soochow University, Suzhou 215006, China}
\affiliation{Institute for Advanced Study, Soochow University, Suzhou 215006, China}
\author{Qing-Feng Sun}
\affiliation{International Center for Quantum Materials, School of Physics, Peking University, Beijing 100871, China}
\affiliation{CAS Center for Excellence in Topological Quantum Computation, University of Chinese Academy of Sciences, Beijing 100190, China}
\affiliation{Hefei National Laboratory, Hefei 230088, China}
\author{X. C. Xie}
\email{xcxie@pku.edu.cn}
\affiliation{International Center for Quantum Materials, School of Physics, Peking University, Beijing 100871, China}
\affiliation{Institute for Nanoelectronic Devices and Quantum Computing, Fudan University, Shanghai 200433, China}
\affiliation{Hefei National Laboratory, Hefei 230088, China}
\begin{abstract}
The quantum Hall effect (QHE), which was observed in 2D electron gas under an external magnetic field, stands out as one of the most remarkable transport phenomena in condensed matter. However, a long standing puzzle remains regarding the observation of the relativistic quantum Hall effect (RQHE). This effect, predicted for a single 2D Dirac cone  immersed in  a magnetic field, is distinguished by the intriguing feature of  half-integer Hall conductivity (HIHC). 
In this work, we demonstrate that the condensed-matter realization of the RQHE and the direct measurement of the HIHC are feasible by investigating the underlying quantum transport mechanism.
We reveal that the manifestation of HIHC is tied to the presence of dissipative half-integer quantized chiral channels circulating along the interface of the RQHE system and a Dirac metal.
Importantly, we find that the Ohmic scaling of the longitudinal conductance of the system plays a key role in  directly measuring the HIHC in experiments.
Furthermore, we propose a feasible experimental scheme based on the 3D topological insulators to  directly measure the HIHC.
Our findings not only uncover the distinct transport mechanism of the HIHC for the RQHE, but also paves the way to the measurement of the HIHC in future experiments.

\end{abstract}

\maketitle

{\emph{Introduction.}}---
The seminal discovery of the quantum Hall effect represents a pivotal milestone in topological states of matter \cite{Klitzing1980,TKNN1982,Hasan2010,Qi2011}. 
The integer quantum Hall effect (IQHE) has been observed in conventional two-dimensional (2D) electron systems subjected to an external magnetic field, where the Hall conductivity takes integer quantized values in units of  $e^2/h$. 
Interestingly, the relativistic quantum Hall effect (RQHE) predicted for a single 2D Dirac fermion possesses half-integer Hall conductivity
(HIHC) $\sigma_{xy}=(\nu +{1}/{2}) e^2/h$ with $\nu$ an integer \cite{Abouelsaood1985, RQHE}.
In retrospect, the initial proposal for realizing RQHE is based on the materials with 2D Dirac-like electronic excitations, such as graphene and 3D topological insulator (TI) surfaces \cite{RQHE,Zheng2002,Gusynin2005,Fu2007PRB,Qi2008,Lee2009}.
However, despite the experimental discoveries of graphene \cite{Novoselov2004} and 3D TIs \cite{Hsieh2009,Roushan2009,YLChen2009}, only an integer Hall conductance has been observed in experiments because the Dirac fermions appear in pairs in these systems \cite{Novoselov2005, ZhangNat2005, Castro2009,Molenkamp2011, Xu2014, Yoshimi2015, Xu2016}. 
Therefore, the direct observation of HIHC remains elusive, and the condensed-matter realization of RQHE is still a significant challenge.


Theoretically, it is widely recognized that the quantization of Hall conductivity for IQHE is intimately linked to the Chern number of the Landau levels \cite{TKNN1982}. 
According to bulk-edge correspondence, Landau levels  undergo deformation along the sample edge, resulting dissipationless chiral edge modes. These dissipationless edge modes, in turn, give rise to the observation of integer Hall conductivity in experiments.
On the other hand, tremendous efforts have also been exerted to uncover the origin of the HIHC in the RQHE.  The HIHC was obtained either by calculating the bulk electromagnetic response of a single 2D Dirac cone or by just dividing integer quantized Hall conductivity of multiple 2D Dirac cones by the number of degenerate Dirac cones \cite{Abouelsaood1985, RQHE,Nomura2008,Qi2008,Sitte2012,Zheng2002,Gusynin2005,Lee2009,Watanabe2010,Zhang_2012}.
However, the microscope edge transport theory on the RQHE remains unknown since the number of chiral edge states cannot be half-integer \cite{Shen2011,Gong2022}.
On a more fundamental level, the transport mechanism of HIHC may extend beyond the realm of dissipationless edge transport for IQHE. 
So far, there are  at least three important issues not answered.
 (i)  Is it really impossible to directly measure the HIHC $\sigma_{xy}=(\nu+1/2)e^2/h$, $\nu\in \mathbb{Z}$  experimentally?
 (ii) Does the chiral edge trajectory exists in the RQHE?
 (iii) What is the transport mechanism of the RQHE?

\begin{figure}[bht]
	\centering
	\includegraphics[width=3.3in]{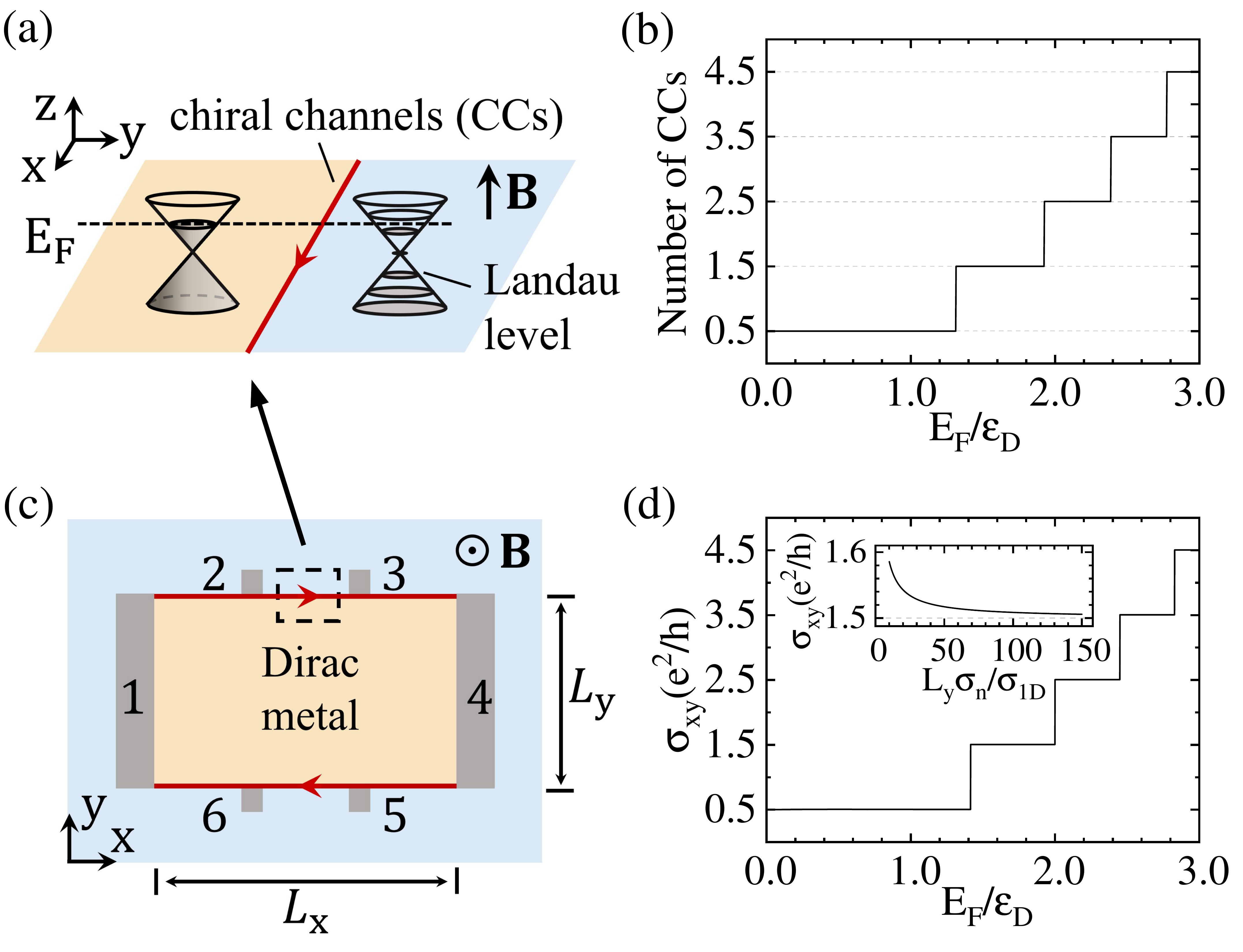}
	\caption{(Color online). 
		(a) Schematic plot of the 2D Dirac fermion system with (without) a perpendicular magnetic field in the blue (yellow) region. The red line indicates chiral channels (CCs). (b) The number of the CCs vesus Fermi energy $E_F/\epsilon_D$.
		(c) Hall-bar measurement of the CCs. 
		(d) The Hall conductivity $\sigma_{xy}$ vesus $E_F/\epsilon_D$ with
		$\sigma_{\text{1D}}/(\sigma_0 L_y)=0.01$. The inset shows $\sigma_{xy}$ as a function of $L_y \sigma_n/\sigma_{\text{1D}}$ with $E_F/\epsilon_D=\sqrt{3}$ and  $\sigma_0= e^2/h $.		
		\label{fig1} }
\end{figure}
In this Letter, we find that the HIHC can be directly measured in experimental setups through investigation of the transport properties of the RQHE.
Inspired by recent experimental observation of the half-quantized anomalous Hall conductance in a hybrid system of massive and massless Dirac cones \cite{Mogi2021}, we examine  a 2D massless Dirac cone under a nonuniform magnetic field.
This setup results in a similar hybrid system, consisting of a 2D massless Dirac metal and the RQHE system, as depicted by the yellow and blue regions in Fig.~\ref{fig1}(a).
We first reveal half-integer quantized chiral channels (CCs) located at the interface [see the red line in Fig.~\ref{fig1}(a)]. 
Then, we establish an analytic relation between the Hall conductivity and the dissipative CCs, and find that the experimental measurement of the HIHC is determined by the Ohmic scaling of the longitudinal conductance.
This relation uncovers bulk-edge correspondence of the RQHE system by elucidating how the half-integer quantized CCs on the edge results in the HIHC within the bulk.
Significantly, to achieve the HIHC in realistic systems, we employ Landauer-B{\"u}ttiker formalism to determine half-integer quantized CCs and HIHC on the surface of the 3D TI, allowing for the direct comparison with experimental observations.
Ultimately, we propose a feasible experimental scheme based on the 3D TI to directly measure the HIHC by a six-terminal device.

{\em Model Hamiltonian and half-integer quantized CCs.}---We begin with the 2D massless Dirac Hamiltonian in a nonuniform magnetic field [see Fig.~\ref{fig1}(a)] that reads:
	\begin{equation}\label{eq1}
		H=\left\{
	\begin{array}{ll}
		-i\hbar v_F  {\bm \nabla}\cdot{\bm \sigma} &(-L_y<y< 0)\\
		v_F(-i\hbar  {\bm \nabla}+e{\bf A})\cdot{\bm \sigma} &(y\geq 0)
	\end{array}\right.
\end{equation}
where $v_F$ denotes the Fermi velocity and the Pauli matrices ${\bm \sigma}$ act in spin space.  For $-L_y< y <0$, the energy $E_{\bf k}=\pm\hbar v_F k$, with the wave vector operator  $(k_x,k_y)=(k\cos \theta,k\sin \theta)$.
For $y \geq 0$, the external magnetic field  $ B_0$ is applied along the $+z$ axis, and the vector potential is taken in the gauge ${\bf A}=-yB_0\hat{e}_x$. Then, the Landau levels forms with energy $\epsilon_n={\rm sgn}(n)\epsilon_D \sqrt{2  \left | n \right |}$ with $n \in \mathbb{Z}$ [see Fig.~\ref{eq1}(a)]. The energy and length scales are set to be $\epsilon_D=\hbar v_F/l_B$ and $l_B=\sqrt{\hbar/(eB_0)}$.
 


To obtain the CCs, we first derive eigenstates by solving the Dirac equation $H\psi_{\bf k}({\bf x})=E_{\bf k}\psi_{\bf k}({\bf x})$. For $-L_y < y < 0$, the eigenstate $\psi_{\bf k}({\bf x})$ reads:
\begin{eqnarray}\label{eq3}
	\psi_{\rm I}({\bf x})=
	\frac{e^{ik_x x}}{\sqrt{4 L_x L_y}}
	\left[
	\begin{pmatrix} 
		e^{-i\frac{\theta}{2}}\\
		e^{i\frac{\theta}{2}} 
	\end{pmatrix} e^{ik_y y}
	+r
	\begin{pmatrix} 
		e^{i\frac{\theta}{2}}\\
		e^{-i\frac{\theta}{2}} 
	\end{pmatrix} e^{-ik_y y}
	\right]
\end{eqnarray}
with a $\theta$-dependent reflection amplitude $r$.
For the magnetic field region ($y \geq 0$), 
\begin{eqnarray}\label{eq4}
	\psi_{\rm II}({\bf x})=
	\frac{t e^{ik_x x}}{\sqrt{4 L_x L_y}}
	\begin{pmatrix} 
		D_{(k l_B)^2/2-1}[\sqrt{2}(y/l_B-k_x l_B)]\\
		-\frac{\sqrt{2}}{k l_B} D_{(k l_B)^2/2}[\sqrt{2}(y/l_B-k_x l_B)] 
	\end{pmatrix}	
\end{eqnarray}
$D_{\nu}(x)$ is the parabolic cylinder function \cite{Gradshteyn1980}. 
$\psi_{\rm II}({\bf x})$ is a decaying wave as $(k l_B)^2/2 \notin \mathbb{Z}$.
By enforcing the continuity of $\psi_{\bf k}({\bf x})$ at $y=0$, we obtain 
\begin{eqnarray}\label{eq6}
	r(k,\theta)
	\equiv e^{i\phi_r(k,\theta)} 
	=
	-\frac{D_2+D_1e^{i\theta}}{D_1 +D_2e^{i\theta}}
\end{eqnarray}
and $t ={-2 i \sin \theta e^{i\theta/2}}/[D_1+D_2e^{i\theta}]$.
Here, we denote $D_1=D_{(k l_B)^2/2-1}(-\sqrt{2} k_x l_B)$ and $D_2=\sqrt{2}/(k l_B) D_{(k l_B)^2/2}(-\sqrt{2} k_x l_B)$.  $\phi_r(k,\theta)$ is the reflection phase, representing the phase shift between the reflected wave and the incident wave. 

Next, we calculate the number of the CCs that directly determines the Hall transport. 
The transverse equilibrium current for occupied states is given by $J_c(E_F)=\lim_{d,L_y\rightarrow\infty}\sum_{E_{\bf k}\leq E_F}\int_{-d}^{+\infty}j_{x,{\bf k}}({\bf x})\mathrm{d}y$ with $d<L_y$, where $j_{x,{\bf k}}({\bf x})=-ev_F \psi_{\bf k}^{\dagger}({\bf x}) \sigma_x \psi_{\bf k}({\bf x})$ is the transverse current density for $\psi_{\bf k}({\bf x})$. 
By taking the partial derivative of $J_c(E_F)$ with $E_F=\hbar v_F k_F$, we obtain
\begin{eqnarray}\label{eq7}
	\frac{h}{e} \frac{\partial J_c(E_F,d)}{\partial E_F}
	=\int_{0}^{\pi} \frac{\mathrm{d} \theta}{2 \pi} \left (-\frac{\partial \phi_r(k_F,\theta)}{\partial \theta} \right )
\end{eqnarray}
which is the total number of CCs. 
The relationship $r(k_F,0)r(k_F,\pi)=-1$ derived from Eq.~(\ref{eq6}) leads to $\phi_r(k_F ,0)-\phi_r(k_F , \pi)=2\pi(n+1/2)$ with $n\in \mathbb{Z}$. 
This ensures the half-integer quantization of CCs [see Fig.~\ref{fig1}(b)] and the final result is given by:
\begin{eqnarray}\label{eq71}
	\frac{h}{e}\frac{\partial J_c(E_F)}{\partial E_F}=  \lfloor ( {E_F}/{\epsilon_D})^2/2 \rfloor+\frac{1}{2} 
\end{eqnarray}
where $\lfloor c \rfloor$ denotes the largest integer less than $c$.  
Note that the equilibrium current and CCs are predominantly located at the interface between the metallic region and the magnetic field region \cite{SM}. This half-integer quantized CCs manifests the bulk-edge correspondence of the RQHE system.


{\em Ohmic Scaling and HIHC.}---
Now, let's establish an analytic relation between the Hall conductivity and the CCs,
and  demonstrate how  Ohmic scaling of conductance determines the existence of the HIHC. 
We consider a metallic system of size $L_x\times L_y$ (yellow region) with half-integer quantized CCs circulating around its edge, as showed in Fig.~\ref{fig1}(c).
To observe a stable HIHC in experiments,  it is essential for both the Hall resistivity and longitudinal resistivity to be independent of system's size.
For this purpose, we assume an Ohmic scaling of the conductance $G_{\text{2D}}$ in the metallic system, i.e. $G_{\text{2D}}=\sigma_nL_y/L_x$, ensuring that the conductivity $\sigma_n$ is unaffected by the system's size.
Furthermore, the dissipation also emerges in the CCs that coexist with the metallic system \cite{SM}.
To incorporate this contribution, we introduce the edge conductivity $\sigma_{\text{1D}}$ that arises from the dissipative CCs, and the edge conductance satisfies 1D Ohmic scaling law, i.e. $G_{\text{1D}}=\sigma_{\text{1D}}/L_x$.

To investigate the transport properties of Hall-bar device in Fig.~\ref{fig1}(c),
a small external bias between leads 1 and 4 is applied to generate a uniform electric field ${\bf E}=(E_x,-E_y)$ \cite{SM}. 
Then, the total current flowing from the lead 1 to 4 is given by $I_x = \sigma_n E_x L_y +\sigma_{\text{1D}}E_x+ e {\partial J_c(E_F)}/{\partial E_F} E_y L_y$. Here the first two terms are the longitudinal current, and the third term is the Hall current from the CCs.
Considering current conservation near the edge $E_y \sigma_n=E_x e {\partial J_c(E_F)}/{\partial E_F}$, the longitudianal and Hall resistivity can be obtained by $\rho_{xx}=E_x/j_x$ and  $\rho_{xy}=E_y/j_x$, with the current density $j_x=I_x/L_y$.
In the Hall-bar measurement, $E_x =( V_2-V_3)/L$ and $E_y=(V_2-V_6)/ L_y$, with $V_i$ the voltage of lead $i$.  $L$ ($L_y$) is the length between the lead 2 and 3 (the lead 2 and 6).
Finally, by using the tensor relation that $\sigma_{xx}=\rho_{xx}/(\rho_{xx}^2+\rho_{xy}^2)$ and $\sigma_{xy}=\rho_{xy}/(\rho_{xx}^2+\rho_{xy}^2)$,
the longitudinal and Hall conductivity are obtained [see more details in Ref.\cite{SM}]:
\begin{eqnarray}\label{eq8}
	\sigma_{xy}(E_F)&=&\left[1+\frac{1}{1+(\sigma_c/\sigma_n)^2}\frac{\sigma_{\text{1D}}}{ \sigma_n L_y }\right] \sigma_c \nonumber \\
	\sigma_{xx}(E_F)&=& \left[1+\frac{1}{1+(\sigma_c/\sigma_n)^2}\frac{\sigma_{\text{1D}}}{\sigma_n L_y}\right]\sigma_n
\end{eqnarray}
where $\sigma_c \equiv e\partial J_c(E_F) / \partial E_F$.
Here, although $G_{\text{2D}}=\sigma_n L_y/L_x$ of the metal is size-independent (when $L_y/L_x$ is fixed),
the longitudinal edge conductance $G_{\text{1D}}=\sigma_{\text{1D}}/L_x$ arising from the CCs is size-dependent.
Therefore, the whole system breaks the 2D Ohmic scaling law, which is the reason why $\sigma_{xy}$ are not quantized in Eq.~(\ref{eq8}).
In Fig.~\ref{fig1}(d), we plot the $\sigma_{xy}$ as a function of the Fermi energy $E_F/\epsilon_D$,  and  assume that $\sigma_n=\sigma_0|E_F|/\epsilon_D$ with $\sigma_0=e^2/h$ since the density of state of 2D massless Dirac cone is proposal to $|E_F|$.
Notably, in the limit that $G_{\text{1D}} \ll G_{\text{2D}}$ (or $\sigma_{\text{1D}} \ll \sigma_n L_y$ ), the Hall conductance in Eq.~(\ref{eq8}) are half-integer quantized to be $\sigma_{xy}=e\partial J_c(E_F) / \partial E_F$ [see also Fig.~\ref{fig1}(d)].
As a result, we conclude that the HIHC is tied to the half-integer quantized CCs
and the Ohmic scaling of the conductance is crucial to the direct observation of the HIHC 
in experiments. 


\begin{figure}[bht]
	\centering
	\includegraphics[width=3.3in]{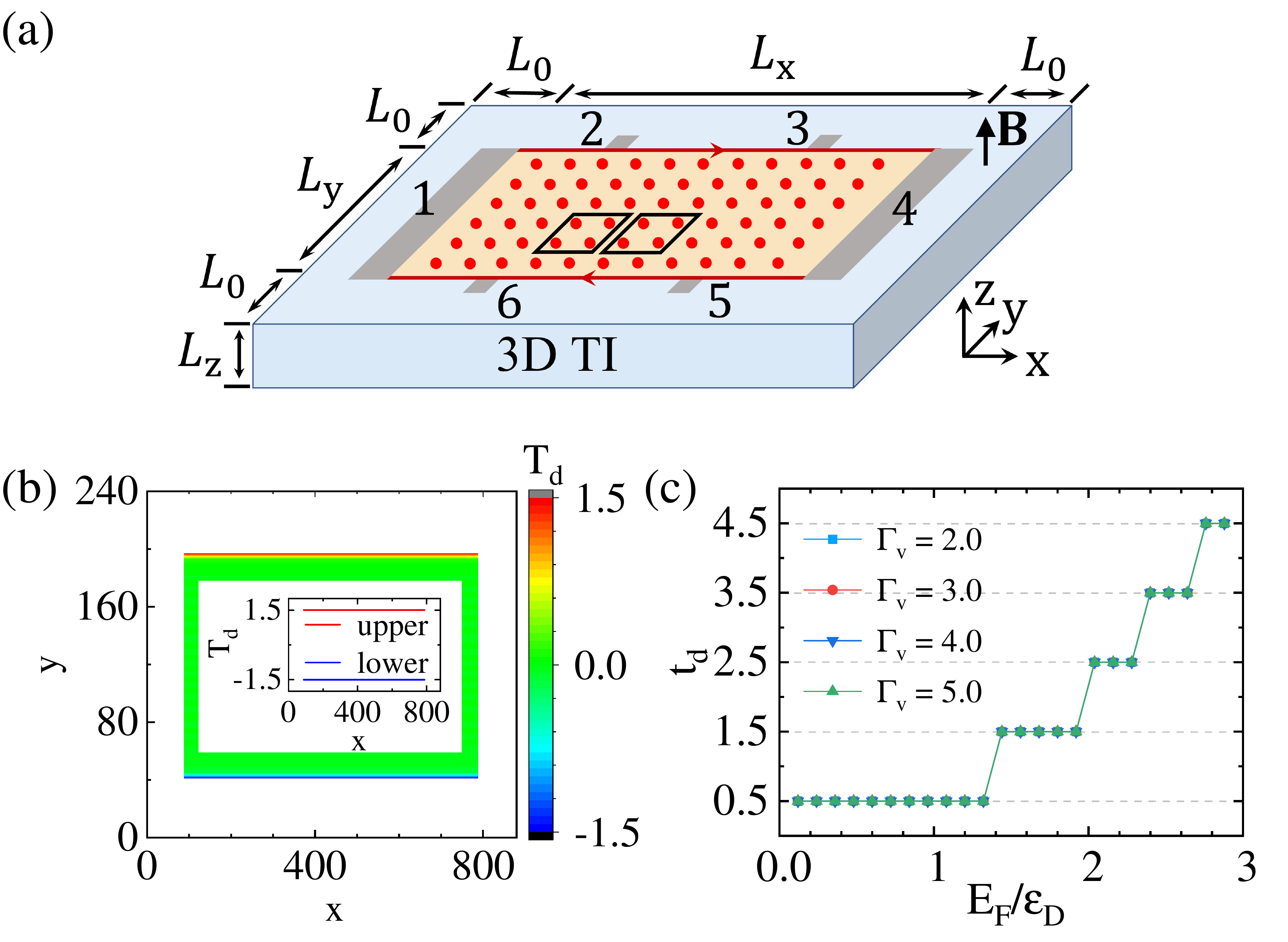}
	\caption{(Color online). 
		(a) Hall-bar measurement of the CCs (red lines) on the 3D TI surface with a perpendicular magnetic field in the blue region. The red balls denotes the sites attached to virtual leads.  
		(b) The spatial distribution of $T_d(x,y)$ with $\Gamma_v=4.0$ and $E_F/\epsilon_D=1.68$. The inset shows $T_d(x,y)$ on the upper ($y=L_0+L_y-r_y+1$) and lower ($y=L_0+1$) interface. (c) $t_d$ versus $E_F$ for different $\Gamma_v$. 
		Here, the lattice constant $a=1$ and the model parameters $A=1.0$, $B=0.5$, $M_0=1$. The magnetic field $B_0=2/201[h/(ea^2)]$ and $\epsilon_D=A/\sqrt{\hbar/(eB_0)}=0.25$.  The size of black boxes in (a) $r_x \times r_y=48\times 48$, $L_x \times L_y \times L_z=800 \times 201 \times 6$, $L_0=40$, $n_x \times n_y=200 \times 51$, and $L=40$ is the length between leads $2$ and $3$.	
		\label{fig2} } 
\end{figure}  

{\em HIHC in the 3D TI.}--To illustrate the feasibility of direct experimental measurement of the HIHC, we employ the Landauer-B{\"u}ttiker formula to numerically calculate the Hall conductivity. 
We consider a 3D TI that hosts a 2D gapless surface Dirac cone \cite{Fu2007PRL,Fu2007PRB} and the effective  four-band TI Hamiltonian is given by \cite{Zhang2009} 
\begin{eqnarray}
	H_{\text{3D}}({\bf k})&=& \sum_{i=x,y,z}Ak_i\sigma_x{\otimes}s_i+(M_0-Bk^2)\sigma_z{\otimes}s_0 \nonumber
\end{eqnarray}
with model parameters $A$, $B$, and $M_0$. $\sigma_i$ and $s_i$ are Pauli matrices for the orbital and spin degrees of freedom, respectively. 
The perpendicular magnetic field is applied in the blue region with ${\bf B}=B_0\hat{e}_z$ (${\bf B}=0$ in the yellow region) [see Fig.~\ref{fig2}(a)].
We discretize the Hamiltonian $H_{\text{3D}}$ into $(L_x+2L_0) \times (L_y+2L_0) \times L_z$ cubic lattice sites.
To obtain the Ohmic scaling of conductance,
we introduce $n_x \times n_y$ B{\"u}ttiker's virtual leads on the yellow region to simulate the dephasing process \cite{SM,Buttiker1986,Buttiker1988,Xing2008,Zhou2022}. This allows the conductance  satisfies the Ohmic scaling when the system size is much larger than the dephasing length. In reality, dephasing process can be caused by electron-electron or electron-phonon interactions \cite{Chakravarty1986,A1985,Stern1990}.
According to Landauer-B{\"u}ttiker formula, the current flowing out from the lead $p$ is
$J_{p}=\frac{e^2}{h}\sum_{q\ne p}\left(T_{qp}V_{p}-T_{pq}V_{q}\right)$
where $V_p$ is the voltage in the lead $p$.  $T_{pq}(E_F)=\mbox{Tr}[{\bm \Gamma}_p{\bm G}^r{\bm \Gamma}_q{\bm G}^a]$ is the transmission coefficient from the lead $q$ to the lead $p$ with the linewidth function ${\bm \Gamma}_p=\Gamma_v {\bf I}_p$ and retarded (advanced) Green's function ${\bf G}^{a(r)}$ \cite{SM}. $\Gamma_v$ is the dephasing strength \cite{Y2008}, and ${\bf I}_p$ is $4n_p \times 4n_p$ unit matrix. $n_p$ is the number of the sites coupling to the lead $p$, where $n_{1(4)}=L_y$ for the real lead 1(4) and $n_p=1$ for other leads.

In order to reveal the half-integer quantized CCs, we investigate the equilibrium current between the virtual leads with the same voltage $V_0=E_F/e$. 
The equilibrium current flowing from the box $a$ to the box $b$ is  $J_{ba}=(e^2/h) T_d(x,y) V_0$, where $T_d(x,y)=\sum_{p\in a,q\in b}\left(T_{qp}-T_{pq}\right)$ and $(x,y)$ is the space coordinate of the virtual lead in the lower right corner of the box $a$. Here, the box $a(b)$ is the left (right) black box with size $r_x \times r_y$ in Fig.~\ref{fig2}(a). 
Figure~\ref{fig2}(b) shows that $T_d$ is half-integer quantized on the upper and lower interface between the metallic region and the magnetic field region, and zero elsewhere. 
For simplicity, we define $t_d\equiv T_d(L_0+L_x/2,L_0+L_y-r_y+1)$ as the value of $T_d$ near the upper interface. In this case, the equilibrium current near the interface is given by $J_c=(e^2/h) t_d V_0$, and $\partial J_c(E_F) / \partial E_F=(e/h) t_d$. 
Figure~\ref{fig2}(c) shows that $t_d=\lfloor(E_F/\epsilon_D)^2/2\rfloor+1/2$ and it is independent of $\Gamma_v$.
To summarize, these findings illustrate the existence of half-integer quantized  CCs circulating around the interface between  the metallic region and magnetic field region on the surface of the 3D TI.

\begin{figure}[thb]
	\centering
	\includegraphics[width=3.3in]{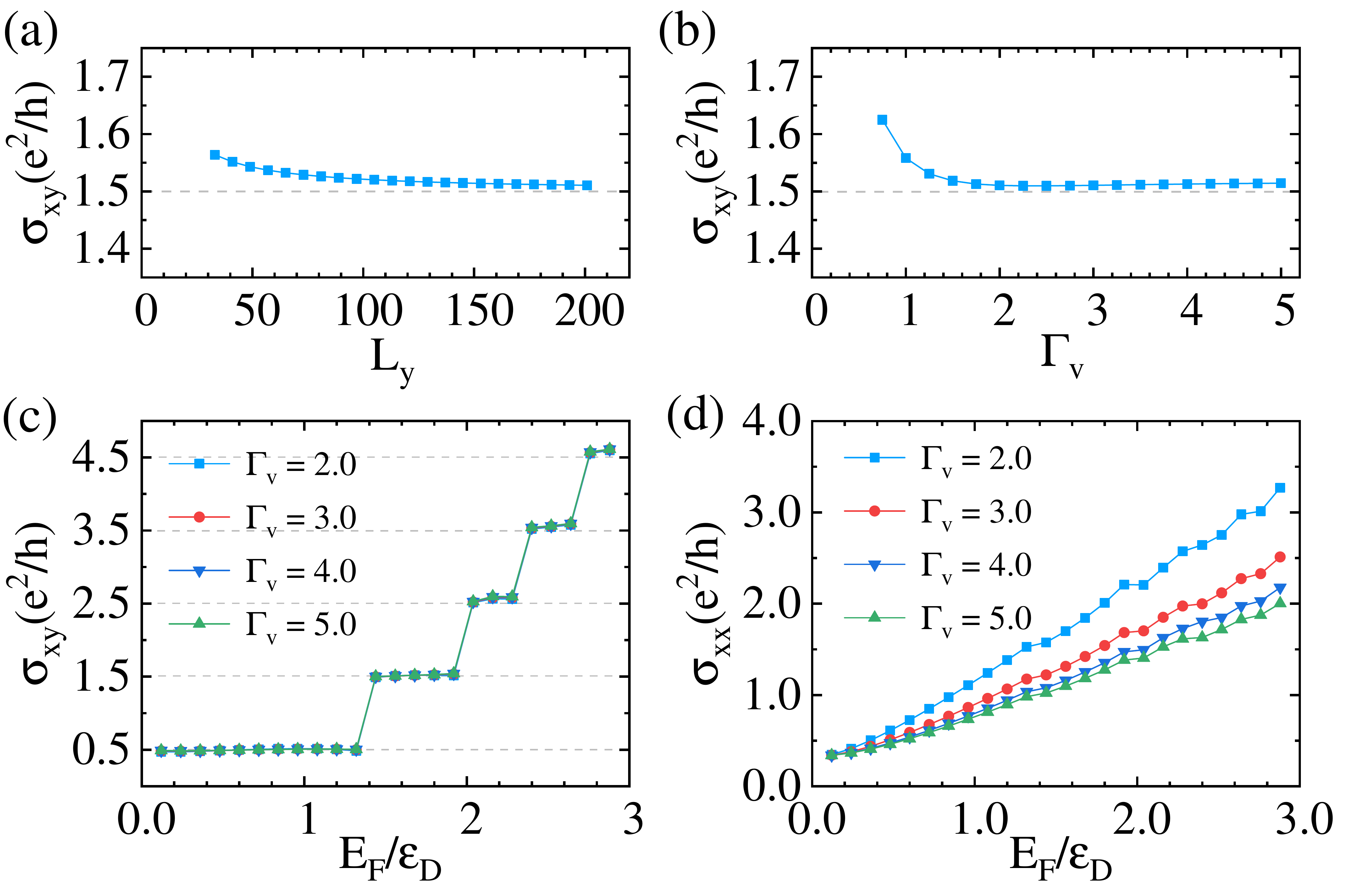}
	\caption{(Color online). 
	$\sigma_{xy}$ versus (a) $L_y$ with  $\Gamma_v=3.0$,
	and (b) $\Gamma_v$ with $L_y=201$. The parameters are $E_F/\epsilon_D=1.68$, $n_y=(L_y-1)/4$, $n_x=4n_y$, $L_x=4n_x$. 
	(c) $\sigma_{xy}$  and (d) $\sigma_{xx}$ as a function of $E_F$ for different $\Gamma_v$.  Other parameters are the same as those in Fig.~\ref{fig2}.
	\label{fig3} }
\end{figure}
Now let's come to investigate the Hall $\sigma_{xy}$ and longitudinal conductivity $\sigma_{xx}$
using the Landauer-B{\"u}ttiker formula  \cite{SM}.
Figures ~\ref{fig3}(a) and (b) show that $\sigma_{xy}$ decreases and converges to $1.5e^2/h$ as $L_y$ increases or as $\sigma_{\text{1D}}$ decreases by increasing $\Gamma_v$, in accordance with Eq.~(\ref{eq8}).
$\sigma_{\text{1D}}$ decrease as $\Gamma_v$ increases because of the dephasing-induced momentum relaxation \cite{Datta2007}. 
Here, $\sigma_{xy}$ is deviated from half-integer quantization because the edge conductance $G_{\text{1D}}$ is size-dependent and breaks the 2D Ohmic scaling.
In large-size limit, $\sigma_{xy}$ is half-integer quantized  in Figs.~\ref{fig3}(c)
agreeing with $\sigma_{xy}=e^2/h(\lfloor(E_F/\epsilon_D)^2/2\rfloor+1/2)$ ,
where $\sigma_{xx}$ varies almost linearly with $E_F$  in Figs.~\ref{fig3}(d).
We stress that the remarkable consistency between the above numerical analysis and the theoretical predictions [Eq.~(\ref{eq8})] strongly demonstrates the reliability of the obtained results.                   
Therefore,  the HIHC can be observed in the 3D TI, where the Ohmic scaling plays a crucial role in the experimental measurement of the HIHC.



\begin{figure}[tbh]
\centering
\includegraphics[width=3.3in]{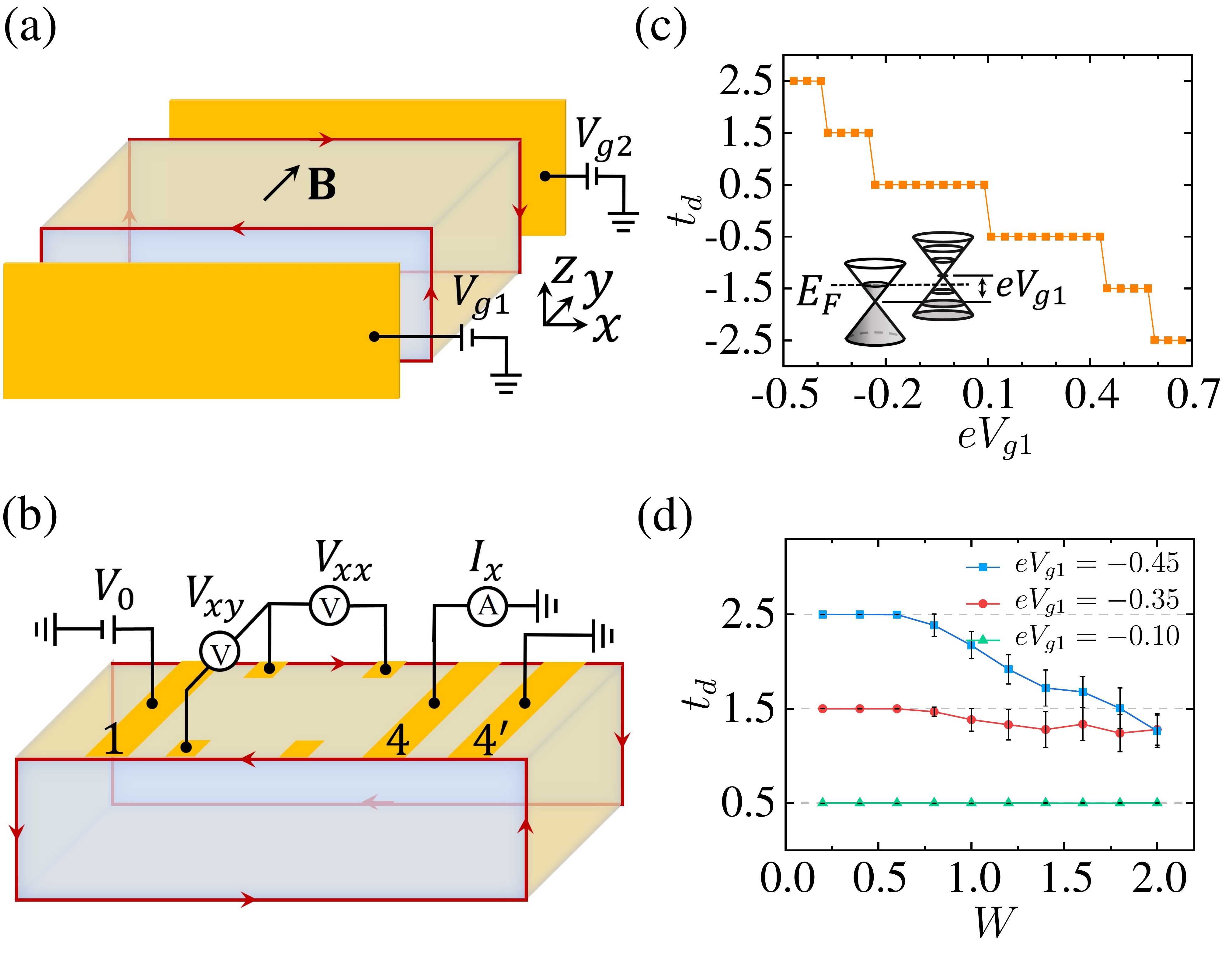}
\caption{(Color online). 
	(a) Schematic plot of a 3D TI in a uniform magnetic field ${\bf B}=B_0\hat{e}_y$ with gate voltages ($V_{g1,2}$) on the front and  back surfaces. 
	The red line represents the CCs. 
	(b)  Schematic plot of the Hall bar device.
	(c) $t_{d}$ versus $eV_{g1}$.
	(d)  $t_{d}$ versus disorder strength $W$ for different $eV_{g1}$.
	Standard deviations for 20 random configurations are shown as the error bars.
     Here, $E_F=0.1$, $B_0=0.01[h/(ea^2)]$, $\Gamma_v=4.0$, $L_x \times L_y \times L_z=140 \times 40 \times 40$,  $n_x \times n_y=140 \times 40$, and $r_x \times r_y =45 \times 10$.
	Other parameters are the same as those in Fig.~\ref{fig2}. 
	\label{fig4} }
\end{figure}
{\em Experimental realization of the HIHC.}--
Due to the experimental challenges in achieving the nonuniform magnetic field in Fig.~\ref{fig2}(a), we propose a more feasible approach to measure the HIHC.
In Fig.~\ref{fig4}(a), a 3D TI is subjected to a uniform magnetic field ${\bf B}=B_0\hat{e}_y$, and a gate voltage $V_{g1}$ ($V_{g2}$) is applied to the front (back) surface to shift the Dirac point by an energy $eV_{g1}$ ($eV_{g2}$).
These two gate voltages will introduce opposite-type (electron-hole) carriers on the front and back surfaces when $eV_{g1}+eV_{g2}=2E_F$ \cite{SM}.
Due to the opposite electromagnetic responses of holes and electrons, the effective magnetic fields on the front and back surfaces are directed inward towards these surfaces. 
In this scenario, other four surfaces act as Dirac metals, encompassed by a pair of anti-parallel CCs on the front and back surfaces. Figure~\ref{fig4}(c) demonstrates that the number of CCs ($t_d$) near the edge of the top surface is quantized in half-integer values.
This quantized CCs is also robust to the weak disorder [see Fig.~\ref{fig4}(d)], where the disorder is introduce as $H_d=V({\bf r})\sigma_0{\otimes}s_0$ and the on-site energy $V({\bf r})$ is uniformly distributed within $[-W/2,W/2]$ with disorder strength $W$. 
Now we can employ the Hall bar device to obtain the HIHC on the top Dirac surface [see Fig.~\ref{fig4}(b)]. Since lead 4 and $4'$ have the same voltage, there is no net current between them. Thus, the current $I_x$ solely arises from the top surface, enabling measurement of the transport properties exclusively on that surface.
From this perspective, the configuration in Fig.~\ref{fig4}(c) is effectively equivalent to the arrangement shown in Fig.~\ref{fig1}(c). Therefore, we can obtain the HIHC in this Hall-bar measurement according to Eq.~(\ref{eq8}). 

{\em Conclusion.}-- In summary, our work confirms the feasibility of directly observing the RQHE and HIHC.
We presents a transport theory of the RQHE, and demonstrates the existence of half-integer quantized CCs at the interface of the RQHE and Dirac metal. We establish a direct relationship between the HIHC and these quantized CCs and uncover the pivotal role of Ohmic scaling in experimentally measuring the HIHC.
Furthermore, we provide a feasible proposal based on the 3D TI to obtain the HIHC and realize the RQHE. 



{\emph{Acknowledgement.}}---  We thank Ming Gong, Wen-Bo Dai, Hailong Li, Runjie Zheng, Dong-Hui Xu for illuminating discussions.
This work was supported by the National Key R\&D Program of China (Grants No. 2022YFA1403700), the National Basic Research Program of China (Grant No. 2015CB921102), 
NSFC (Grants No. 11921005, and No. 11974256), 
the Innovation Program for Quantum Science and Technology (Grant No. 2021ZD0302403), and the Strategic Priority Research Program of Chinese Academy of Sciences (Grant No. XDB28000000).
C.-Z.C. is also funded by the Priority Academic Program Development of Jiangsu Higher Education Institutions.

\bibliographystyle{apsrev4-2} 
\bibliography{ref}

\begin{thebibliography}{42}%
\makeatletter
\providecommand \@ifxundefined [1]{%
 \@ifx{#1\undefined}
}%
\providecommand \@ifnum [1]{%
 \ifnum #1\expandafter \@firstoftwo
 \else \expandafter \@secondoftwo
 \fi
}%
\providecommand \@ifx [1]{%
 \ifx #1\expandafter \@firstoftwo
 \else \expandafter \@secondoftwo
 \fi
}%
\providecommand \natexlab [1]{#1}%
\providecommand \enquote  [1]{``#1''}%
\providecommand \bibnamefont  [1]{#1}%
\providecommand \bibfnamefont [1]{#1}%
\providecommand \citenamefont [1]{#1}%
\providecommand \href@noop [0]{\@secondoftwo}%
\providecommand \href [0]{\begingroup \@sanitize@url \@href}%
\providecommand \@href[1]{\@@startlink{#1}\@@href}%
\providecommand \@@href[1]{\endgroup#1\@@endlink}%
\providecommand \@sanitize@url [0]{\catcode `\\12\catcode `\$12\catcode
  `\&12\catcode `\#12\catcode `\^12\catcode `\_12\catcode `\%12\relax}%
\providecommand \@@startlink[1]{}%
\providecommand \@@endlink[0]{}%
\providecommand \url  [0]{\begingroup\@sanitize@url \@url }%
\providecommand \@url [1]{\endgroup\@href {#1}{\urlprefix }}%
\providecommand \urlprefix  [0]{URL }%
\providecommand \Eprint [0]{\href }%
\providecommand \doibase [0]{http://dx.doi.org/}%
\providecommand \selectlanguage [0]{\@gobble}%
\providecommand \bibinfo  [0]{\@secondoftwo}%
\providecommand \bibfield  [0]{\@secondoftwo}%
\providecommand \translation [1]{[#1]}%
\providecommand \BibitemOpen [0]{}%
\providecommand \bibitemStop [0]{}%
\providecommand \bibitemNoStop [0]{.\EOS\space}%
\providecommand \EOS [0]{\spacefactor3000\relax}%
\providecommand \BibitemShut  [1]{\csname bibitem#1\endcsname}%
\let\auto@bib@innerbib\@empty
\bibitem [{\citenamefont {Klitzing}\ \emph {et~al.}(1980)\citenamefont
  {Klitzing}, \citenamefont {Dorda},\ and\ \citenamefont
  {Pepper}}]{Klitzing1980}%
  \BibitemOpen
  \bibfield  {author} {\bibinfo {author} {\bibfnamefont {K.~v.}\ \bibnamefont
  {Klitzing}}, \bibinfo {author} {\bibfnamefont {G.}~\bibnamefont {Dorda}}, \
  and\ \bibinfo {author} {\bibfnamefont {M.}~\bibnamefont {Pepper}},\ }\href
  {\doibase 10.1103/PhysRevLett.45.494} {\bibfield  {journal} {\bibinfo
  {journal} {Phys. Rev. Lett.}\ }\textbf {\bibinfo {volume} {45}},\ \bibinfo
  {pages} {494} (\bibinfo {year} {1980})}\BibitemShut {NoStop}%
\bibitem [{\citenamefont {Thouless}\ \emph {et~al.}(1982)\citenamefont
  {Thouless}, \citenamefont {Kohmoto}, \citenamefont {Nightingale},\ and\
  \citenamefont {den Nijs}}]{TKNN1982}%
  \BibitemOpen
  \bibfield  {author} {\bibinfo {author} {\bibfnamefont {D.~J.}\ \bibnamefont
  {Thouless}}, \bibinfo {author} {\bibfnamefont {M.}~\bibnamefont {Kohmoto}},
  \bibinfo {author} {\bibfnamefont {M.~P.}\ \bibnamefont {Nightingale}}, \ and\
  \bibinfo {author} {\bibfnamefont {M.}~\bibnamefont {den Nijs}},\ }\href
  {\doibase 10.1103/PhysRevLett.49.405} {\bibfield  {journal} {\bibinfo
  {journal} {Phys. Rev. Lett.}\ }\textbf {\bibinfo {volume} {49}},\ \bibinfo
  {pages} {405} (\bibinfo {year} {1982})}\BibitemShut {NoStop}%
\bibitem [{\citenamefont {Hasan}\ and\ \citenamefont {Kane}(2010)}]{Hasan2010}%
  \BibitemOpen
  \bibfield  {author} {\bibinfo {author} {\bibfnamefont {M.~Z.}\ \bibnamefont
  {Hasan}}\ and\ \bibinfo {author} {\bibfnamefont {C.~L.}\ \bibnamefont
  {Kane}},\ }\href {\doibase 10.1103/RevModPhys.82.3045} {\bibfield  {journal}
  {\bibinfo  {journal} {Rev. Mod. Phys.}\ }\textbf {\bibinfo {volume} {82}},\
  \bibinfo {pages} {3045} (\bibinfo {year} {2010})}\BibitemShut {NoStop}%
\bibitem [{\citenamefont {Qi}\ and\ \citenamefont {Zhang}(2011)}]{Qi2011}%
  \BibitemOpen
  \bibfield  {author} {\bibinfo {author} {\bibfnamefont {X.-L.}\ \bibnamefont
  {Qi}}\ and\ \bibinfo {author} {\bibfnamefont {S.-C.}\ \bibnamefont {Zhang}},\
  }\href {\doibase 10.1103/RevModPhys.83.1057} {\bibfield  {journal} {\bibinfo
  {journal} {Rev. Mod. Phys.}\ }\textbf {\bibinfo {volume} {83}},\ \bibinfo
  {pages} {1057} (\bibinfo {year} {2011})}\BibitemShut {NoStop}%
\bibitem [{\citenamefont {Abouelsaood}(1985)}]{Abouelsaood1985}%
  \BibitemOpen
  \bibfield  {author} {\bibinfo {author} {\bibfnamefont {A.}~\bibnamefont
  {Abouelsaood}},\ }\href {\doibase 10.1103/PhysRevLett.54.1973} {\bibfield
  {journal} {\bibinfo  {journal} {Phys. Rev. Lett.}\ }\textbf {\bibinfo
  {volume} {54}},\ \bibinfo {pages} {1973} (\bibinfo {year}
  {1985})}\BibitemShut {NoStop}%
\bibitem [{\citenamefont {Schakel}(1991)}]{RQHE}%
  \BibitemOpen
  \bibfield  {author} {\bibinfo {author} {\bibfnamefont {A.~M.~J.}\
  \bibnamefont {Schakel}},\ }\href {\doibase 10.1103/PhysRevD.43.1428}
  {\bibfield  {journal} {\bibinfo  {journal} {Phys. Rev. D}\ }\textbf {\bibinfo
  {volume} {43}},\ \bibinfo {pages} {1428} (\bibinfo {year}
  {1991})}\BibitemShut {NoStop}%
\bibitem [{\citenamefont {Zheng}\ and\ \citenamefont {Ando}(2002)}]{Zheng2002}%
  \BibitemOpen
  \bibfield  {author} {\bibinfo {author} {\bibfnamefont {Y.}~\bibnamefont
  {Zheng}}\ and\ \bibinfo {author} {\bibfnamefont {T.}~\bibnamefont {Ando}},\
  }\href {\doibase 10.1103/PhysRevB.65.245420} {\bibfield  {journal} {\bibinfo
  {journal} {Phys. Rev. B}\ }\textbf {\bibinfo {volume} {65}},\ \bibinfo
  {pages} {245420} (\bibinfo {year} {2002})}\BibitemShut {NoStop}%
\bibitem [{\citenamefont {Gusynin}\ and\ \citenamefont
  {Sharapov}(2005)}]{Gusynin2005}%
  \BibitemOpen
  \bibfield  {author} {\bibinfo {author} {\bibfnamefont {V.~P.}\ \bibnamefont
  {Gusynin}}\ and\ \bibinfo {author} {\bibfnamefont {S.~G.}\ \bibnamefont
  {Sharapov}},\ }\href {\doibase 10.1103/PhysRevLett.95.146801} {\bibfield
  {journal} {\bibinfo  {journal} {Phys. Rev. Lett.}\ }\textbf {\bibinfo
  {volume} {95}},\ \bibinfo {pages} {146801} (\bibinfo {year}
  {2005})}\BibitemShut {NoStop}%
\bibitem [{\citenamefont {Fu}\ and\ \citenamefont {Kane}(2007)}]{Fu2007PRB}%
  \BibitemOpen
  \bibfield  {author} {\bibinfo {author} {\bibfnamefont {L.}~\bibnamefont
  {Fu}}\ and\ \bibinfo {author} {\bibfnamefont {C.~L.}\ \bibnamefont {Kane}},\
  }\href {\doibase 10.1103/PhysRevB.76.045302} {\bibfield  {journal} {\bibinfo
  {journal} {Phys. Rev. B}\ }\textbf {\bibinfo {volume} {76}},\ \bibinfo
  {pages} {045302} (\bibinfo {year} {2007})}\BibitemShut {NoStop}%
\bibitem [{\citenamefont {Qi}\ \emph {et~al.}(2008)\citenamefont {Qi},
  \citenamefont {Hughes},\ and\ \citenamefont {Zhang}}]{Qi2008}%
  \BibitemOpen
  \bibfield  {author} {\bibinfo {author} {\bibfnamefont {X.-L.}\ \bibnamefont
  {Qi}}, \bibinfo {author} {\bibfnamefont {T.~L.}\ \bibnamefont {Hughes}}, \
  and\ \bibinfo {author} {\bibfnamefont {S.-C.}\ \bibnamefont {Zhang}},\ }\href
  {\doibase 10.1103/PhysRevB.78.195424} {\bibfield  {journal} {\bibinfo
  {journal} {Phys. Rev. B}\ }\textbf {\bibinfo {volume} {78}},\ \bibinfo
  {pages} {195424} (\bibinfo {year} {2008})}\BibitemShut {NoStop}%
\bibitem [{\citenamefont {Lee}(2009)}]{Lee2009}%
  \BibitemOpen
  \bibfield  {author} {\bibinfo {author} {\bibfnamefont {D.-H.}\ \bibnamefont
  {Lee}},\ }\href {\doibase 10.1103/PhysRevLett.103.196804} {\bibfield
  {journal} {\bibinfo  {journal} {Phys. Rev. Lett.}\ }\textbf {\bibinfo
  {volume} {103}},\ \bibinfo {pages} {196804} (\bibinfo {year}
  {2009})}\BibitemShut {NoStop}%
\bibitem [{\citenamefont {Novoselov}\ \emph {et~al.}(2004)\citenamefont
  {Novoselov}, \citenamefont {Geim}, \citenamefont {Morozov}, \citenamefont
  {Jiang}, \citenamefont {Zhang}, \citenamefont {Dubonos}, \citenamefont
  {Grigorieva},\ and\ \citenamefont {Firsov}}]{Novoselov2004}%
  \BibitemOpen
  \bibfield  {author} {\bibinfo {author} {\bibfnamefont {K.~S.}\ \bibnamefont
  {Novoselov}}, \bibinfo {author} {\bibfnamefont {A.~K.}\ \bibnamefont {Geim}},
  \bibinfo {author} {\bibfnamefont {S.~V.}\ \bibnamefont {Morozov}}, \bibinfo
  {author} {\bibfnamefont {D.}~\bibnamefont {Jiang}}, \bibinfo {author}
  {\bibfnamefont {Y.}~\bibnamefont {Zhang}}, \bibinfo {author} {\bibfnamefont
  {S.~V.}\ \bibnamefont {Dubonos}}, \bibinfo {author} {\bibfnamefont {I.~V.}\
  \bibnamefont {Grigorieva}}, \ and\ \bibinfo {author} {\bibfnamefont {A.~A.}\
  \bibnamefont {Firsov}},\ }\href {\doibase 10.1126/science.1102896} {\bibfield
   {journal} {\bibinfo  {journal} {Science}\ }\textbf {\bibinfo {volume}
  {306}},\ \bibinfo {pages} {666} (\bibinfo {year} {2004})}\BibitemShut
  {NoStop}%
\bibitem [{\citenamefont {Hsieh}\ \emph {et~al.}(2009)\citenamefont {Hsieh},
  \citenamefont {Xia}, \citenamefont {Qian}, \citenamefont {Wray},
  \citenamefont {Dil}, \citenamefont {Meier}, \citenamefont {Osterwalder},
  \citenamefont {Patthey}, \citenamefont {Checkelsky}, \citenamefont {Ong},
  \citenamefont {Fedorov}, \citenamefont {Lin}, \citenamefont {Bansil},
  \citenamefont {Grauer}, \citenamefont {Hor}, \citenamefont {Cava},\ and\
  \citenamefont {Hasan}}]{Hsieh2009}%
  \BibitemOpen
  \bibfield  {author} {\bibinfo {author} {\bibfnamefont {D.}~\bibnamefont
  {Hsieh}}, \bibinfo {author} {\bibfnamefont {Y.}~\bibnamefont {Xia}}, \bibinfo
  {author} {\bibfnamefont {D.}~\bibnamefont {Qian}}, \bibinfo {author}
  {\bibfnamefont {L.}~\bibnamefont {Wray}}, \bibinfo {author} {\bibfnamefont
  {J.~H.}\ \bibnamefont {Dil}}, \bibinfo {author} {\bibfnamefont
  {F.}~\bibnamefont {Meier}}, \bibinfo {author} {\bibfnamefont
  {J.}~\bibnamefont {Osterwalder}}, \bibinfo {author} {\bibfnamefont
  {L.}~\bibnamefont {Patthey}}, \bibinfo {author} {\bibfnamefont {J.~G.}\
  \bibnamefont {Checkelsky}}, \bibinfo {author} {\bibfnamefont {N.~P.}\
  \bibnamefont {Ong}}, \bibinfo {author} {\bibfnamefont {A.~V.}\ \bibnamefont
  {Fedorov}}, \bibinfo {author} {\bibfnamefont {H.}~\bibnamefont {Lin}},
  \bibinfo {author} {\bibfnamefont {A.}~\bibnamefont {Bansil}}, \bibinfo
  {author} {\bibfnamefont {D.}~\bibnamefont {Grauer}}, \bibinfo {author}
  {\bibfnamefont {Y.~S.}\ \bibnamefont {Hor}}, \bibinfo {author} {\bibfnamefont
  {R.~J.}\ \bibnamefont {Cava}}, \ and\ \bibinfo {author} {\bibfnamefont
  {M.~Z.}\ \bibnamefont {Hasan}},\ }\href {\doibase 10.1038/nature08234}
  {\bibfield  {journal} {\bibinfo  {journal} {Nature}\ }\textbf {\bibinfo
  {volume} {460}},\ \bibinfo {pages} {1101} (\bibinfo {year}
  {2009})}\BibitemShut {NoStop}%
\bibitem [{\citenamefont {Roushan}\ \emph {et~al.}(2009)\citenamefont
  {Roushan}, \citenamefont {Seo}, \citenamefont {Parker}, \citenamefont {Hor},
  \citenamefont {Hsieh}, \citenamefont {Qian}, \citenamefont {Richardella},
  \citenamefont {Hasan}, \citenamefont {Cava},\ and\ \citenamefont
  {Yazdani}}]{Roushan2009}%
  \BibitemOpen
  \bibfield  {author} {\bibinfo {author} {\bibfnamefont {P.}~\bibnamefont
  {Roushan}}, \bibinfo {author} {\bibfnamefont {J.}~\bibnamefont {Seo}},
  \bibinfo {author} {\bibfnamefont {C.~V.}\ \bibnamefont {Parker}}, \bibinfo
  {author} {\bibfnamefont {Y.~S.}\ \bibnamefont {Hor}}, \bibinfo {author}
  {\bibfnamefont {D.}~\bibnamefont {Hsieh}}, \bibinfo {author} {\bibfnamefont
  {D.}~\bibnamefont {Qian}}, \bibinfo {author} {\bibfnamefont {A.}~\bibnamefont
  {Richardella}}, \bibinfo {author} {\bibfnamefont {M.~Z.}\ \bibnamefont
  {Hasan}}, \bibinfo {author} {\bibfnamefont {R.~J.}\ \bibnamefont {Cava}}, \
  and\ \bibinfo {author} {\bibfnamefont {A.}~\bibnamefont {Yazdani}},\ }\href
  {\doibase 10.1038/nature08308} {\bibfield  {journal} {\bibinfo  {journal}
  {Nature}\ }\textbf {\bibinfo {volume} {460}},\ \bibinfo {pages} {1106}
  (\bibinfo {year} {2009})}\BibitemShut {NoStop}%
\bibitem [{\citenamefont {Chen}\ \emph {et~al.}(2009)\citenamefont {Chen},
  \citenamefont {Analytis}, \citenamefont {Chu}, \citenamefont {Liu},
  \citenamefont {Mo}, \citenamefont {Qi}, \citenamefont {Zhang}, \citenamefont
  {Lu}, \citenamefont {Dai}, \citenamefont {Fang}, \citenamefont {Zhang},
  \citenamefont {Fisher}, \citenamefont {Hussain},\ and\ \citenamefont
  {Shen}}]{YLChen2009}%
  \BibitemOpen
  \bibfield  {author} {\bibinfo {author} {\bibfnamefont {Y.~L.}\ \bibnamefont
  {Chen}}, \bibinfo {author} {\bibfnamefont {J.~G.}\ \bibnamefont {Analytis}},
  \bibinfo {author} {\bibfnamefont {J.-H.}\ \bibnamefont {Chu}}, \bibinfo
  {author} {\bibfnamefont {Z.~K.}\ \bibnamefont {Liu}}, \bibinfo {author}
  {\bibfnamefont {S.-K.}\ \bibnamefont {Mo}}, \bibinfo {author} {\bibfnamefont
  {X.~L.}\ \bibnamefont {Qi}}, \bibinfo {author} {\bibfnamefont {H.~J.}\
  \bibnamefont {Zhang}}, \bibinfo {author} {\bibfnamefont {D.~H.}\ \bibnamefont
  {Lu}}, \bibinfo {author} {\bibfnamefont {X.}~\bibnamefont {Dai}}, \bibinfo
  {author} {\bibfnamefont {Z.}~\bibnamefont {Fang}}, \bibinfo {author}
  {\bibfnamefont {S.~C.}\ \bibnamefont {Zhang}}, \bibinfo {author}
  {\bibfnamefont {I.~R.}\ \bibnamefont {Fisher}}, \bibinfo {author}
  {\bibfnamefont {Z.}~\bibnamefont {Hussain}}, \ and\ \bibinfo {author}
  {\bibfnamefont {Z.-X.}\ \bibnamefont {Shen}},\ }\href {\doibase
  10.1126/science.1173034} {\bibfield  {journal} {\bibinfo  {journal}
  {Science}\ }\textbf {\bibinfo {volume} {325}},\ \bibinfo {pages} {178}
  (\bibinfo {year} {2009})}\BibitemShut {NoStop}%
\bibitem [{\citenamefont {Novoselov}\ \emph {et~al.}(2005)\citenamefont
  {Novoselov}, \citenamefont {Geim}, \citenamefont {Morozov}, \citenamefont
  {Jiang}, \citenamefont {Katsnelson}, \citenamefont {Grigorieva},
  \citenamefont {Dubonos},\ and\ \citenamefont {Firsov}}]{Novoselov2005}%
  \BibitemOpen
  \bibfield  {author} {\bibinfo {author} {\bibfnamefont {K.~S.}\ \bibnamefont
  {Novoselov}}, \bibinfo {author} {\bibfnamefont {A.~K.}\ \bibnamefont {Geim}},
  \bibinfo {author} {\bibfnamefont {S.~V.}\ \bibnamefont {Morozov}}, \bibinfo
  {author} {\bibfnamefont {D.}~\bibnamefont {Jiang}}, \bibinfo {author}
  {\bibfnamefont {M.~I.}\ \bibnamefont {Katsnelson}}, \bibinfo {author}
  {\bibfnamefont {I.~V.}\ \bibnamefont {Grigorieva}}, \bibinfo {author}
  {\bibfnamefont {S.~V.}\ \bibnamefont {Dubonos}}, \ and\ \bibinfo {author}
  {\bibfnamefont {A.~A.}\ \bibnamefont {Firsov}},\ }\href {\doibase
  10.1038/nature04233} {\bibfield  {journal} {\bibinfo  {journal} {Nature}\
  }\textbf {\bibinfo {volume} {438}},\ \bibinfo {pages} {197} (\bibinfo {year}
  {2005})}\BibitemShut {NoStop}%
\bibitem [{\citenamefont {Zhang}\ \emph {et~al.}(2005)\citenamefont {Zhang},
  \citenamefont {Tan}, \citenamefont {Stormer},\ and\ \citenamefont
  {Kim}}]{ZhangNat2005}%
  \BibitemOpen
  \bibfield  {author} {\bibinfo {author} {\bibfnamefont {Y.}~\bibnamefont
  {Zhang}}, \bibinfo {author} {\bibfnamefont {Y.-W.}\ \bibnamefont {Tan}},
  \bibinfo {author} {\bibfnamefont {H.~L.}\ \bibnamefont {Stormer}}, \ and\
  \bibinfo {author} {\bibfnamefont {P.}~\bibnamefont {Kim}},\ }\href {\doibase
  10.1038/nature04235} {\bibfield  {journal} {\bibinfo  {journal} {Nature}\
  }\textbf {\bibinfo {volume} {438}},\ \bibinfo {pages} {201} (\bibinfo {year}
  {2005})}\BibitemShut {NoStop}%
\bibitem [{\citenamefont {Castro~Neto}\ \emph {et~al.}(2009)\citenamefont
  {Castro~Neto}, \citenamefont {Guinea}, \citenamefont {Peres}, \citenamefont
  {Novoselov},\ and\ \citenamefont {Geim}}]{Castro2009}%
  \BibitemOpen
  \bibfield  {author} {\bibinfo {author} {\bibfnamefont {A.~H.}\ \bibnamefont
  {Castro~Neto}}, \bibinfo {author} {\bibfnamefont {F.}~\bibnamefont {Guinea}},
  \bibinfo {author} {\bibfnamefont {N.~M.~R.}\ \bibnamefont {Peres}}, \bibinfo
  {author} {\bibfnamefont {K.~S.}\ \bibnamefont {Novoselov}}, \ and\ \bibinfo
  {author} {\bibfnamefont {A.~K.}\ \bibnamefont {Geim}},\ }\href {\doibase
  10.1103/RevModPhys.81.109} {\bibfield  {journal} {\bibinfo  {journal} {Rev.
  Mod. Phys.}\ }\textbf {\bibinfo {volume} {81}},\ \bibinfo {pages} {109}
  (\bibinfo {year} {2009})}\BibitemShut {NoStop}%
\bibitem [{\citenamefont {Br\"une}\ \emph {et~al.}(2011)\citenamefont
  {Br\"une}, \citenamefont {Liu}, \citenamefont {Novik}, \citenamefont
  {Hankiewicz}, \citenamefont {Buhmann}, \citenamefont {Chen}, \citenamefont
  {Qi}, \citenamefont {Shen}, \citenamefont {Zhang},\ and\ \citenamefont
  {Molenkamp}}]{Molenkamp2011}%
  \BibitemOpen
  \bibfield  {author} {\bibinfo {author} {\bibfnamefont {C.}~\bibnamefont
  {Br\"une}}, \bibinfo {author} {\bibfnamefont {C.~X.}\ \bibnamefont {Liu}},
  \bibinfo {author} {\bibfnamefont {E.~G.}\ \bibnamefont {Novik}}, \bibinfo
  {author} {\bibfnamefont {E.~M.}\ \bibnamefont {Hankiewicz}}, \bibinfo
  {author} {\bibfnamefont {H.}~\bibnamefont {Buhmann}}, \bibinfo {author}
  {\bibfnamefont {Y.~L.}\ \bibnamefont {Chen}}, \bibinfo {author}
  {\bibfnamefont {X.~L.}\ \bibnamefont {Qi}}, \bibinfo {author} {\bibfnamefont
  {Z.~X.}\ \bibnamefont {Shen}}, \bibinfo {author} {\bibfnamefont {S.~C.}\
  \bibnamefont {Zhang}}, \ and\ \bibinfo {author} {\bibfnamefont {L.~W.}\
  \bibnamefont {Molenkamp}},\ }\href {\doibase 10.1103/PhysRevLett.106.126803}
  {\bibfield  {journal} {\bibinfo  {journal} {Phys. Rev. Lett.}\ }\textbf
  {\bibinfo {volume} {106}},\ \bibinfo {pages} {126803} (\bibinfo {year}
  {2011})}\BibitemShut {NoStop}%
\bibitem [{\citenamefont {Xu}\ \emph {et~al.}(2014)\citenamefont {Xu},
  \citenamefont {Miotkowski}, \citenamefont {Liu}, \citenamefont {Tian},
  \citenamefont {Nam}, \citenamefont {Alidoust}, \citenamefont {Hu},
  \citenamefont {Shih}, \citenamefont {Hasan},\ and\ \citenamefont
  {Chen}}]{Xu2014}%
  \BibitemOpen
  \bibfield  {author} {\bibinfo {author} {\bibfnamefont {Y.}~\bibnamefont
  {Xu}}, \bibinfo {author} {\bibfnamefont {I.}~\bibnamefont {Miotkowski}},
  \bibinfo {author} {\bibfnamefont {C.}~\bibnamefont {Liu}}, \bibinfo {author}
  {\bibfnamefont {J.}~\bibnamefont {Tian}}, \bibinfo {author} {\bibfnamefont
  {H.}~\bibnamefont {Nam}}, \bibinfo {author} {\bibfnamefont {N.}~\bibnamefont
  {Alidoust}}, \bibinfo {author} {\bibfnamefont {J.}~\bibnamefont {Hu}},
  \bibinfo {author} {\bibfnamefont {C.~K.}\ \bibnamefont {Shih}}, \bibinfo
  {author} {\bibfnamefont {M.~Z.}\ \bibnamefont {Hasan}}, \ and\ \bibinfo
  {author} {\bibfnamefont {Y.~P.}\ \bibnamefont {Chen}},\ }\href
  {https://doi.org/10.1038/nphys3140} {\bibfield  {journal} {\bibinfo
  {journal} {Nature Physics}\ }\textbf {\bibinfo {volume} {10}},\ \bibinfo
  {pages} {956} (\bibinfo {year} {2014})}\BibitemShut {NoStop}%
\bibitem [{\citenamefont {Yoshimi}\ \emph {et~al.}(2015)\citenamefont
  {Yoshimi}, \citenamefont {Yasuda}, \citenamefont {Tsukazaki}, \citenamefont
  {Takahashi}, \citenamefont {Nagaosa}, \citenamefont {Kawasaki},\ and\
  \citenamefont {Tokura}}]{Yoshimi2015}%
  \BibitemOpen
  \bibfield  {author} {\bibinfo {author} {\bibfnamefont {R.}~\bibnamefont
  {Yoshimi}}, \bibinfo {author} {\bibfnamefont {K.}~\bibnamefont {Yasuda}},
  \bibinfo {author} {\bibfnamefont {A.}~\bibnamefont {Tsukazaki}}, \bibinfo
  {author} {\bibfnamefont {K.}~\bibnamefont {Takahashi}}, \bibinfo {author}
  {\bibfnamefont {N.}~\bibnamefont {Nagaosa}}, \bibinfo {author} {\bibfnamefont
  {M.}~\bibnamefont {Kawasaki}}, \ and\ \bibinfo {author} {\bibfnamefont
  {Y.}~\bibnamefont {Tokura}},\ }\href {https://doi.org/10.1038/ncomms9530}
  {\bibfield  {journal} {\bibinfo  {journal} {Nature communications}\ }\textbf
  {\bibinfo {volume} {6}},\ \bibinfo {pages} {8530} (\bibinfo {year}
  {2015})}\BibitemShut {NoStop}%
\bibitem [{\citenamefont {Xu}\ \emph {et~al.}(2016)\citenamefont {Xu},
  \citenamefont {Miotkowski},\ and\ \citenamefont {Chen}}]{Xu2016}%
  \BibitemOpen
  \bibfield  {author} {\bibinfo {author} {\bibfnamefont {Y.}~\bibnamefont
  {Xu}}, \bibinfo {author} {\bibfnamefont {I.}~\bibnamefont {Miotkowski}}, \
  and\ \bibinfo {author} {\bibfnamefont {Y.~P.}\ \bibnamefont {Chen}},\ }\href
  {https://doi.org/10.1038/ncomms11434} {\bibfield  {journal} {\bibinfo
  {journal} {Nature Communications}\ }\textbf {\bibinfo {volume} {7}},\
  \bibinfo {pages} {11434} (\bibinfo {year} {2016})}\BibitemShut {NoStop}%
\bibitem [{\citenamefont {Nomura}\ \emph {et~al.}(2008)\citenamefont {Nomura},
  \citenamefont {Ryu}, \citenamefont {Koshino}, \citenamefont {Mudry},\ and\
  \citenamefont {Furusaki}}]{Nomura2008}%
  \BibitemOpen
  \bibfield  {author} {\bibinfo {author} {\bibfnamefont {K.}~\bibnamefont
  {Nomura}}, \bibinfo {author} {\bibfnamefont {S.}~\bibnamefont {Ryu}},
  \bibinfo {author} {\bibfnamefont {M.}~\bibnamefont {Koshino}}, \bibinfo
  {author} {\bibfnamefont {C.}~\bibnamefont {Mudry}}, \ and\ \bibinfo {author}
  {\bibfnamefont {A.}~\bibnamefont {Furusaki}},\ }\href {\doibase
  10.1103/PhysRevLett.100.246806} {\bibfield  {journal} {\bibinfo  {journal}
  {Phys. Rev. Lett.}\ }\textbf {\bibinfo {volume} {100}},\ \bibinfo {pages}
  {246806} (\bibinfo {year} {2008})}\BibitemShut {NoStop}%
\bibitem [{\citenamefont {Sitte}\ \emph {et~al.}(2012)\citenamefont {Sitte},
  \citenamefont {Rosch}, \citenamefont {Altman},\ and\ \citenamefont
  {Fritz}}]{Sitte2012}%
  \BibitemOpen
  \bibfield  {author} {\bibinfo {author} {\bibfnamefont {M.}~\bibnamefont
  {Sitte}}, \bibinfo {author} {\bibfnamefont {A.}~\bibnamefont {Rosch}},
  \bibinfo {author} {\bibfnamefont {E.}~\bibnamefont {Altman}}, \ and\ \bibinfo
  {author} {\bibfnamefont {L.}~\bibnamefont {Fritz}},\ }\href {\doibase
  10.1103/PhysRevLett.108.126807} {\bibfield  {journal} {\bibinfo  {journal}
  {Phys. Rev. Lett.}\ }\textbf {\bibinfo {volume} {108}},\ \bibinfo {pages}
  {126807} (\bibinfo {year} {2012})}\BibitemShut {NoStop}%
\bibitem [{\citenamefont {Watanabe}\ \emph {et~al.}(2010)\citenamefont
  {Watanabe}, \citenamefont {Hatsugai},\ and\ \citenamefont
  {Aoki}}]{Watanabe2010}%
  \BibitemOpen
  \bibfield  {author} {\bibinfo {author} {\bibfnamefont {H.}~\bibnamefont
  {Watanabe}}, \bibinfo {author} {\bibfnamefont {Y.}~\bibnamefont {Hatsugai}},
  \ and\ \bibinfo {author} {\bibfnamefont {H.}~\bibnamefont {Aoki}},\ }\href
  {\doibase 10.1103/PhysRevB.82.241403} {\bibfield  {journal} {\bibinfo
  {journal} {Phys. Rev. B}\ }\textbf {\bibinfo {volume} {82}},\ \bibinfo
  {pages} {241403} (\bibinfo {year} {2010})}\BibitemShut {NoStop}%
\bibitem [{\citenamefont {Zhang}\ \emph {et~al.}(2011)\citenamefont {Zhang},
  \citenamefont {Wang},\ and\ \citenamefont {Xie}}]{Zhang_2012}%
  \BibitemOpen
  \bibfield  {author} {\bibinfo {author} {\bibfnamefont {Y.-Y.}\ \bibnamefont
  {Zhang}}, \bibinfo {author} {\bibfnamefont {X.-R.}\ \bibnamefont {Wang}}, \
  and\ \bibinfo {author} {\bibfnamefont {X.~C.}\ \bibnamefont {Xie}},\ }\href
  {\doibase 10.1088/0953-8984/24/1/015004} {\bibfield  {journal} {\bibinfo
  {journal} {Journal of Physics: Condensed Matter}\ }\textbf {\bibinfo {volume}
  {24}},\ \bibinfo {pages} {015004} (\bibinfo {year} {2011})}\BibitemShut
  {NoStop}%
\bibitem [{\citenamefont {Chu}\ \emph {et~al.}(2011)\citenamefont {Chu},
  \citenamefont {Shi},\ and\ \citenamefont {Shen}}]{Shen2011}%
  \BibitemOpen
  \bibfield  {author} {\bibinfo {author} {\bibfnamefont {R.-L.}\ \bibnamefont
  {Chu}}, \bibinfo {author} {\bibfnamefont {J.}~\bibnamefont {Shi}}, \ and\
  \bibinfo {author} {\bibfnamefont {S.-Q.}\ \bibnamefont {Shen}},\ }\href
  {\doibase 10.1103/PhysRevB.84.085312} {\bibfield  {journal} {\bibinfo
  {journal} {Phys. Rev. B}\ }\textbf {\bibinfo {volume} {84}},\ \bibinfo
  {pages} {085312} (\bibinfo {year} {2011})}\BibitemShut {NoStop}%
\bibitem [{\citenamefont {Gong}\ \emph {et~al.}(2023)\citenamefont {Gong},
  \citenamefont {Liu}, \citenamefont {Jiang}, \citenamefont {Chen},\ and\
  \citenamefont {Xie}}]{Gong2022}%
  \BibitemOpen
  \bibfield  {author} {\bibinfo {author} {\bibfnamefont {M.}~\bibnamefont
  {Gong}}, \bibinfo {author} {\bibfnamefont {H.}~\bibnamefont {Liu}}, \bibinfo
  {author} {\bibfnamefont {H.}~\bibnamefont {Jiang}}, \bibinfo {author}
  {\bibfnamefont {C.-Z.}\ \bibnamefont {Chen}}, \ and\ \bibinfo {author}
  {\bibfnamefont {X.~C.}\ \bibnamefont {Xie}},\ }\href {\doibase
  10.1093/nsr/nwad025} {\bibfield  {journal} {\bibinfo  {journal} {National
  Science Review}\ ,\ \bibinfo {pages} {nwad025}} (\bibinfo {year}
  {2023})}\BibitemShut {NoStop}%
\bibitem [{\citenamefont {{Mogi}}\ \emph {et~al.}(2022)\citenamefont {{Mogi}},
  \citenamefont {{Okamura}}, \citenamefont {{Kawamura}}, \citenamefont
  {{Yoshimi}}, \citenamefont {{Yasuda}}, \citenamefont {{Tsukazaki}},
  \citenamefont {{Takahashi}}, \citenamefont {{Morimoto}}, \citenamefont
  {{Nagaosa}}, \citenamefont {{Kawasaki}}, \citenamefont {{Takahashi}},\ and\
  \citenamefont {{Tokura}}}]{Mogi2021}%
  \BibitemOpen
  \bibfield  {author} {\bibinfo {author} {\bibfnamefont {M.}~\bibnamefont
  {{Mogi}}}, \bibinfo {author} {\bibfnamefont {Y.}~\bibnamefont {{Okamura}}},
  \bibinfo {author} {\bibfnamefont {M.}~\bibnamefont {{Kawamura}}}, \bibinfo
  {author} {\bibfnamefont {R.}~\bibnamefont {{Yoshimi}}}, \bibinfo {author}
  {\bibfnamefont {K.}~\bibnamefont {{Yasuda}}}, \bibinfo {author}
  {\bibfnamefont {A.}~\bibnamefont {{Tsukazaki}}}, \bibinfo {author}
  {\bibfnamefont {K.~S.}\ \bibnamefont {{Takahashi}}}, \bibinfo {author}
  {\bibfnamefont {T.}~\bibnamefont {{Morimoto}}}, \bibinfo {author}
  {\bibfnamefont {N.}~\bibnamefont {{Nagaosa}}}, \bibinfo {author}
  {\bibfnamefont {M.}~\bibnamefont {{Kawasaki}}}, \bibinfo {author}
  {\bibfnamefont {Y.}~\bibnamefont {{Takahashi}}}, \ and\ \bibinfo {author}
  {\bibfnamefont {Y.}~\bibnamefont {{Tokura}}},\ }\href
  {https://doi.org/10.1038/s41567-021-01490-y} {\bibfield  {journal} {\bibinfo
  {journal} {Nature Physics}\ } (\bibinfo {year} {2022})}\BibitemShut {NoStop}%
\bibitem [{\citenamefont {Gradshteyn}\ and\ \citenamefont
  {Ryzhik}(1980)}]{Gradshteyn1980}%
  \BibitemOpen
  \bibfield  {author} {\bibinfo {author} {\bibfnamefont {I.~S.}\ \bibnamefont
  {Gradshteyn}}\ and\ \bibinfo {author} {\bibfnamefont {I.~M.}\ \bibnamefont
  {Ryzhik}},\ }\href@noop {} {\emph {\bibinfo {title} {Table of Integrals,
  Series, and Product}}}\ (\bibinfo  {publisher} {Academic Press},\ \bibinfo
  {year} {1980})\BibitemShut {NoStop}%
\bibitem [{SM()}]{SM}%
  \BibitemOpen
  \href@noop {} {\enquote {\bibinfo {title} {See supplemental material for
  detailed discussions},}\ }\BibitemShut {NoStop}%
\bibitem [{\citenamefont {Fu}\ \emph {et~al.}(2007)\citenamefont {Fu},
  \citenamefont {Kane},\ and\ \citenamefont {Mele}}]{Fu2007PRL}%
  \BibitemOpen
  \bibfield  {author} {\bibinfo {author} {\bibfnamefont {L.}~\bibnamefont
  {Fu}}, \bibinfo {author} {\bibfnamefont {C.~L.}\ \bibnamefont {Kane}}, \ and\
  \bibinfo {author} {\bibfnamefont {E.~J.}\ \bibnamefont {Mele}},\ }\href
  {\doibase 10.1103/PhysRevLett.98.106803} {\bibfield  {journal} {\bibinfo
  {journal} {Phys. Rev. Lett.}\ }\textbf {\bibinfo {volume} {98}},\ \bibinfo
  {pages} {106803} (\bibinfo {year} {2007})}\BibitemShut {NoStop}%
\bibitem [{\citenamefont {Zhang}\ \emph {et~al.}(2009)\citenamefont {Zhang},
  \citenamefont {Liu}, \citenamefont {Qi}, \citenamefont {Dai}, \citenamefont
  {Fang},\ and\ \citenamefont {Zhang}}]{Zhang2009}%
  \BibitemOpen
  \bibfield  {author} {\bibinfo {author} {\bibfnamefont {H.}~\bibnamefont
  {Zhang}}, \bibinfo {author} {\bibfnamefont {C.-X.}\ \bibnamefont {Liu}},
  \bibinfo {author} {\bibfnamefont {X.-L.}\ \bibnamefont {Qi}}, \bibinfo
  {author} {\bibfnamefont {X.}~\bibnamefont {Dai}}, \bibinfo {author}
  {\bibfnamefont {Z.}~\bibnamefont {Fang}}, \ and\ \bibinfo {author}
  {\bibfnamefont {S.-C.}\ \bibnamefont {Zhang}},\ }\href
  {https://doi.org/10.1038/nphys1270} {\bibfield  {journal} {\bibinfo
  {journal} {Nature physics}\ }\textbf {\bibinfo {volume} {5}},\ \bibinfo
  {pages} {438} (\bibinfo {year} {2009})}\BibitemShut {NoStop}%
\bibitem [{\citenamefont {B\"uttiker}(1986)}]{Buttiker1986}%
  \BibitemOpen
  \bibfield  {author} {\bibinfo {author} {\bibfnamefont {M.}~\bibnamefont
  {B\"uttiker}},\ }\href {\doibase 10.1103/PhysRevLett.57.1761} {\bibfield
  {journal} {\bibinfo  {journal} {Phys. Rev. Lett.}\ }\textbf {\bibinfo
  {volume} {57}},\ \bibinfo {pages} {1761} (\bibinfo {year}
  {1986})}\BibitemShut {NoStop}%
\bibitem [{\citenamefont {B\"uttiker}(1988)}]{Buttiker1988}%
  \BibitemOpen
  \bibfield  {author} {\bibinfo {author} {\bibfnamefont {M.}~\bibnamefont
  {B\"uttiker}},\ }\href {\doibase 10.1147/rd.323.0317} {\bibfield  {journal}
  {\bibinfo  {journal} {IBM Journal of Research and Development}\ }\textbf
  {\bibinfo {volume} {32}},\ \bibinfo {pages} {317} (\bibinfo {year}
  {1988})}\BibitemShut {NoStop}%
\bibitem [{\citenamefont {Xing}\ \emph
  {et~al.}(2008{\natexlab{a}})\citenamefont {Xing}, \citenamefont {Sun},\ and\
  \citenamefont {Wang}}]{Xing2008}%
  \BibitemOpen
  \bibfield  {author} {\bibinfo {author} {\bibfnamefont {Y.}~\bibnamefont
  {Xing}}, \bibinfo {author} {\bibfnamefont {Q.-f.}\ \bibnamefont {Sun}}, \
  and\ \bibinfo {author} {\bibfnamefont {J.}~\bibnamefont {Wang}},\ }\href
  {\doibase 10.1103/PhysRevB.77.115346} {\bibfield  {journal} {\bibinfo
  {journal} {Phys. Rev. B}\ }\textbf {\bibinfo {volume} {77}},\ \bibinfo
  {pages} {115346} (\bibinfo {year} {2008}{\natexlab{a}})}\BibitemShut
  {NoStop}%
\bibitem [{\citenamefont {Zhou}\ \emph {et~al.}(2022)\citenamefont {Zhou},
  \citenamefont {Li}, \citenamefont {Xu}, \citenamefont {Chen}, \citenamefont
  {Sun},\ and\ \citenamefont {Xie}}]{Zhou2022}%
  \BibitemOpen
  \bibfield  {author} {\bibinfo {author} {\bibfnamefont {H.}~\bibnamefont
  {Zhou}}, \bibinfo {author} {\bibfnamefont {H.}~\bibnamefont {Li}}, \bibinfo
  {author} {\bibfnamefont {D.-H.}\ \bibnamefont {Xu}}, \bibinfo {author}
  {\bibfnamefont {C.-Z.}\ \bibnamefont {Chen}}, \bibinfo {author}
  {\bibfnamefont {Q.-F.}\ \bibnamefont {Sun}}, \ and\ \bibinfo {author}
  {\bibfnamefont {X.~C.}\ \bibnamefont {Xie}},\ }\href {\doibase
  10.1103/PhysRevLett.129.096601} {\bibfield  {journal} {\bibinfo  {journal}
  {Phys. Rev. Lett.}\ }\textbf {\bibinfo {volume} {129}},\ \bibinfo {pages}
  {096601} (\bibinfo {year} {2022})}\BibitemShut {NoStop}%
\bibitem [{\citenamefont {Chakravarty}\ and\ \citenamefont
  {Schmid}(1986)}]{Chakravarty1986}%
  \BibitemOpen
  \bibfield  {author} {\bibinfo {author} {\bibfnamefont {S.}~\bibnamefont
  {Chakravarty}}\ and\ \bibinfo {author} {\bibfnamefont {A.}~\bibnamefont
  {Schmid}},\ }\href {\doibase https://doi.org/10.1016/0370-1573(86)90027-X}
  {\bibfield  {journal} {\bibinfo  {journal} {Physics Reports}\ }\textbf
  {\bibinfo {volume} {140}},\ \bibinfo {pages} {193} (\bibinfo {year}
  {1986})}\BibitemShut {NoStop}%
\bibitem [{\citenamefont {Altshuler}\ and\ \citenamefont
  {Aronov}(1985)}]{A1985}%
  \BibitemOpen
  \bibfield  {author} {\bibinfo {author} {\bibfnamefont {B.~L.}\ \bibnamefont
  {Altshuler}}\ and\ \bibinfo {author} {\bibfnamefont {A.~G.}\ \bibnamefont
  {Aronov}},\ }\href@noop {} {\emph {\bibinfo {title} {Electron--electron
  interaction in disordered conductors}}},\ Vol.~\bibinfo {volume} {10}\
  (\bibinfo  {publisher} {Elsevier},\ \bibinfo {year} {1985})\BibitemShut
  {NoStop}%
\bibitem [{\citenamefont {Stern}\ \emph {et~al.}(1990)\citenamefont {Stern},
  \citenamefont {Aharonov},\ and\ \citenamefont {Imry}}]{Stern1990}%
  \BibitemOpen
  \bibfield  {author} {\bibinfo {author} {\bibfnamefont {A.}~\bibnamefont
  {Stern}}, \bibinfo {author} {\bibfnamefont {Y.}~\bibnamefont {Aharonov}}, \
  and\ \bibinfo {author} {\bibfnamefont {Y.}~\bibnamefont {Imry}},\ }\href
  {\doibase 10.1103/PhysRevA.41.3436} {\bibfield  {journal} {\bibinfo
  {journal} {Phys. Rev. A}\ }\textbf {\bibinfo {volume} {41}},\ \bibinfo
  {pages} {3436} (\bibinfo {year} {1990})}\BibitemShut {NoStop}%
\bibitem [{\citenamefont {Xing}\ \emph
  {et~al.}(2008{\natexlab{b}})\citenamefont {Xing}, \citenamefont {Sun},\ and\
  \citenamefont {Wang}}]{Y2008}%
  \BibitemOpen
  \bibfield  {author} {\bibinfo {author} {\bibfnamefont {Y.}~\bibnamefont
  {Xing}}, \bibinfo {author} {\bibfnamefont {Q.-f.}\ \bibnamefont {Sun}}, \
  and\ \bibinfo {author} {\bibfnamefont {J.}~\bibnamefont {Wang}},\ }\href
  {\doibase 10.1103/PhysRevB.77.115346} {\bibfield  {journal} {\bibinfo
  {journal} {Phys. Rev. B}\ }\textbf {\bibinfo {volume} {77}},\ \bibinfo
  {pages} {115346} (\bibinfo {year} {2008}{\natexlab{b}})}\BibitemShut
  {NoStop}%
\bibitem [{\citenamefont {Golizadeh-Mojarad}\ and\ \citenamefont
  {Datta}(2007)}]{Datta2007}%
  \BibitemOpen
  \bibfield  {author} {\bibinfo {author} {\bibfnamefont {R.}~\bibnamefont
  {Golizadeh-Mojarad}}\ and\ \bibinfo {author} {\bibfnamefont {S.}~\bibnamefont
  {Datta}},\ }\href {\doibase 10.1103/PhysRevB.75.081301} {\bibfield  {journal}
  {\bibinfo  {journal} {Phys. Rev. B}\ }\textbf {\bibinfo {volume} {75}},\
  \bibinfo {pages} {081301} (\bibinfo {year} {2007})}\BibitemShut {NoStop}%
\end{thebibliography}%


\begin{thebibliography}{5}%
\makeatletter
\providecommand \@ifxundefined [1]{%
 \@ifx{#1\undefined}
}%
\providecommand \@ifnum [1]{%
 \ifnum #1\expandafter \@firstoftwo
 \else \expandafter \@secondoftwo
 \fi
}%
\providecommand \@ifx [1]{%
 \ifx #1\expandafter \@firstoftwo
 \else \expandafter \@secondoftwo
 \fi
}%
\providecommand \natexlab [1]{#1}%
\providecommand \enquote  [1]{``#1''}%
\providecommand \bibnamefont  [1]{#1}%
\providecommand \bibfnamefont [1]{#1}%
\providecommand \citenamefont [1]{#1}%
\providecommand \href@noop [0]{\@secondoftwo}%
\providecommand \href [0]{\begingroup \@sanitize@url \@href}%
\providecommand \@href[1]{\@@startlink{#1}\@@href}%
\providecommand \@@href[1]{\endgroup#1\@@endlink}%
\providecommand \@sanitize@url [0]{\catcode `\\12\catcode `\$12\catcode
  `\&12\catcode `\#12\catcode `\^12\catcode `\_12\catcode `\%12\relax}%
\providecommand \@@startlink[1]{}%
\providecommand \@@endlink[0]{}%
\providecommand \url  [0]{\begingroup\@sanitize@url \@url }%
\providecommand \@url [1]{\endgroup\@href {#1}{\urlprefix }}%
\providecommand \urlprefix  [0]{URL }%
\providecommand \Eprint [0]{\href }%
\providecommand \doibase [0]{http://dx.doi.org/}%
\providecommand \selectlanguage [0]{\@gobble}%
\providecommand \bibinfo  [0]{\@secondoftwo}%
\providecommand \bibfield  [0]{\@secondoftwo}%
\providecommand \translation [1]{[#1]}%
\providecommand \BibitemOpen [0]{}%
\providecommand \bibitemStop [0]{}%
\providecommand \bibitemNoStop [0]{.\EOS\space}%
\providecommand \EOS [0]{\spacefactor3000\relax}%
\providecommand \BibitemShut  [1]{\csname bibitem#1\endcsname}%
\let\auto@bib@innerbib\@empty
\bibitem [{\citenamefont {Gradshteyn}\ and\ \citenamefont
  {Ryzhik}(1980)}]{Gradshteyn1980}%
  \BibitemOpen
  \bibfield  {author} {\bibinfo {author} {\bibfnamefont {I.~S.}\ \bibnamefont
  {Gradshteyn}}\ and\ \bibinfo {author} {\bibfnamefont {I.~M.}\ \bibnamefont
  {Ryzhik}},\ }\href@noop {} {\emph {\bibinfo {title} {Table of Integrals,
  Series, and Product}}}\ (\bibinfo  {publisher} {Academic Press},\ \bibinfo
  {year} {1980})\BibitemShut {NoStop}%
\bibitem [{\citenamefont {Hofstadter}(1976)}]{Hofstadter1976}%
  \BibitemOpen
  \bibfield  {author} {\bibinfo {author} {\bibfnamefont {D.~R.}\ \bibnamefont
  {Hofstadter}},\ }\href {\doibase 10.1103/PhysRevB.14.2239} {\bibfield
  {journal} {\bibinfo  {journal} {Phys. Rev. B}\ }\textbf {\bibinfo {volume}
  {14}},\ \bibinfo {pages} {2239} (\bibinfo {year} {1976})}\BibitemShut
  {NoStop}%
\bibitem [{\citenamefont {Xing}\ \emph {et~al.}(2008)\citenamefont {Xing},
  \citenamefont {Sun},\ and\ \citenamefont {Wang}}]{Y2008}%
  \BibitemOpen
  \bibfield  {author} {\bibinfo {author} {\bibfnamefont {Y.}~\bibnamefont
  {Xing}}, \bibinfo {author} {\bibfnamefont {Q.-f.}\ \bibnamefont {Sun}}, \
  and\ \bibinfo {author} {\bibfnamefont {J.}~\bibnamefont {Wang}},\ }\href
  {\doibase 10.1103/PhysRevB.77.115346} {\bibfield  {journal} {\bibinfo
  {journal} {Phys. Rev. B}\ }\textbf {\bibinfo {volume} {77}},\ \bibinfo
  {pages} {115346} (\bibinfo {year} {2008})}\BibitemShut {NoStop}%
\bibitem [{\citenamefont {B\"uttiker}(1988)}]{Buttiker1988}%
  \BibitemOpen
  \bibfield  {author} {\bibinfo {author} {\bibfnamefont {M.}~\bibnamefont
  {B\"uttiker}},\ }\href {\doibase 10.1147/rd.323.0317} {\bibfield  {journal}
  {\bibinfo  {journal} {IBM Journal of Research and Development}\ }\textbf
  {\bibinfo {volume} {32}},\ \bibinfo {pages} {317} (\bibinfo {year}
  {1988})}\BibitemShut {NoStop}%
\bibitem [{\citenamefont {Datta}(1995)}]{Data1995}%
  \BibitemOpen
  \bibfield  {author} {\bibinfo {author} {\bibfnamefont {S.}~\bibnamefont
  {Datta}},\ }\href@noop {} {\emph {\bibinfo {title} {Electronic Transport in
  Mesoscopic Systems}}}\ (\bibinfo  {publisher} {Cambridge University Press,
  Cambridge},\ \bibinfo {year} {1995})\BibitemShut {NoStop}%
\end{thebibliography}%
\end{document}


\title{Supplementary Materials for ``Dissipative Chiral Channels, Ohmic Scaling and Half-integer Hall Conductivity from the Relativistic Quantum Hall Effect''}
\author{Humian Zhou}
\affiliation{International Center for Quantum Materials, School of Physics, Peking University, Beijing 100871, China}
\author{Chui-Zhen Chen}
\email{czchen@suda.edu.cn}
\affiliation{School of Physical Science and Technology, Soochow University, Suzhou 215006, China}
\affiliation{Institute for Advanced Study, Soochow University, Suzhou 215006, China}
\author{Qing-Feng Sun}
\affiliation{International Center for Quantum Materials, School of Physics, Peking University, Beijing 100871, China}
\affiliation{CAS Center for Excellence in Topological Quantum Computation, University of Chinese Academy of Sciences, Beijing 100190, China}
\affiliation{Hefei National Laboratory, Hefei 230088, China}
\author{X. C. Xie}
\email{xcxie@pku.edu.cn}
\affiliation{International Center for Quantum Materials, School of Physics, Peking University, Beijing 100871, China}
\affiliation{Institute for Nanoelectronic Devices and Quantum Computing, Fudan University, Shanghai 200433, China}
\affiliation{Hefei National Laboratory, Hefei 230088, China}
\date{\today }

\maketitle
\tableofcontents
\section{The eigenstates and eigenvalues}

In this section, we will derive the eigenstates and eigenvalues of $H$, where $H=v_F(-i\hbar  {\bm \nabla}+e{\bf A})\cdot{\bm \sigma}$. The perpendicular magnetic field is given by ${\bf B}=B(y)\hat{e}_z$ with $B(y)=0$ within the strip $-L_y< y<0$ and $B(y)=B_0$ elsewhere. We choose the vector potential in the gauge ${\bf A}=A(y)\hat{e}_x$ with $B(y)=-\partial_y A(y)$:
\begin{eqnarray}\label{eq1}
	A(y)=  
	\begin{cases}
		-yB_0,&y \geq 0 \\
		0,&-L_y< y<0 \\
		-(y+L_y)B_0,&y\leq-L_y \\
	\end{cases}	.
\end{eqnarray}
The eigenstate $\psi_{\bf k}({\bf x})$ and eigenvalue $E_{\bf k}$ satisfy the equation $H\psi_{\bf k}({\bf x})=E_{\bf k}\psi_{\bf k}({\bf x})$. Here, ${\bf k}=(k_x,k_y)=(k\cos \theta,k\sin \theta)$ is the wave vector, and $\theta$ is the incidence angle. The Hamiltonian $H$ is translationally  invariant along the $x$ direction, so $k_x$ is conserved. We define $\psi_{\bf k}({\bf x})=(\psi_+,\psi_{-})^\mathrm{T} e^{i k_x x}$, and obtain the coupled equation:
\begin{eqnarray}\label{eq2}
	[\pm \partial_y + k_x + (e/\hbar) A(y) ]\psi_{\pm}=k \psi_{\mp}
	,
\end{eqnarray}
where $k=E_{\bf k}/(\hbar v_F)$. We then decouple the Eq.~\ref{eq2} as:
\begin{eqnarray}\label{eq3}
	[\partial^2_y \pm (e/\hbar) \partial_y  A(y)-(k_x+(e/\hbar) A(y))^2 + k^2 ]\psi_{\pm}=0
	.
\end{eqnarray}
The solution of the Eq.~\ref{eq3} in the three regions is as follows. For $-L_y< y<0$, the Eq.~\ref{eq2} becomes 
\begin{eqnarray}\label{eq4}
	[\partial^2_y -k_x^2 + k^2 ]\psi_{\pm}=0
	.
\end{eqnarray}
 The expression of $\psi_{\bf k}({\bf x})$ in this region is 
\begin{eqnarray}\label{eq5}
	\psi_{\rm I}({\bf x})=
	\frac{e^{ik_x x}}{\sqrt{4 L_x L_y}}
	\left[
	\begin{pmatrix} 
		e^{-i\frac{\theta}{2}}\\
		e^{i\frac{\theta}{2}} 
	\end{pmatrix} e^{ik_y y}
	+r
	\begin{pmatrix} 
		e^{i\frac{\theta}{2}}\\
		e^{-i\frac{\theta}{2}} 
	\end{pmatrix} e^{-ik_y y}
	\right]
	,
\end{eqnarray}
which is the superposition of an incident plane wave and a reflected plane wave with $\theta$-dependent reflection amplitude $r$.
Here, $k_x=2\pi n/L_x $  with $n \in \mathbb{Z}$ since we apply the periodic boundary condition in the $x$ direction. 

In the region $y \geq 0$, the Eq.~\ref{eq2} becomes 
\begin{eqnarray}\label{eq6}
	[l_B^2\partial^2_y \mp 1-(y/l_B-l_B k_x)^2 + (k l_B)^2 ]\psi_{\pm}=0
	,
\end{eqnarray}
where $l_B=\sqrt{\hbar/(eB_0)}$ is the magnetic length.
Considering the eigenstate is finite at $y=+\infty$, the expression of $\psi_{\bf k}({\bf x})$ in this region is 
\begin{eqnarray}\label{eq7}
	\psi_{\rm II}({\bf x})=
	\frac{t e^{ik_x x}}{\sqrt{4 L_x L_y}}
	\begin{pmatrix} 
		D_{(k l_B)^2/2-1}[\sqrt{2}(y/l_B-k_x l_B)]\\
		-\frac{\sqrt{2}}{k l_B} D_{(k l_B)^2/2}[\sqrt{2}(y/l_B-k_x l_B)] 
	\end{pmatrix}
    ,	
\end{eqnarray}
where $D_{\nu}(x)$ is the parabolic cylinder function \cite{Gradshteyn1980}. 
Here, $t$ is a complex coefficient.
For a given uniform perpendicular magnetic field, 
the Landau levels forms with energy $\epsilon_n={\rm sgn}(n)\epsilon_D \sqrt{2  \left | n \right |}$ with $n \in \mathbb{Z}$. 
$\epsilon_D=\hbar v_F/l_B$ is an energy scale.
$\psi_{\rm II}({\bf x})$ is a decaying wave as $(k l_B)^2/2 \notin \mathbb{Z}$.

Similarly, for $x\leq-L_y$, 
\begin{eqnarray}\label{eq8}
	[l_B^2\partial^2_y \mp 1-((y+L_y)/l_B-l_B k_x)^2 + (k l_B)^2 ]\psi_{\pm}=0,
\end{eqnarray}
and $\psi_{\bf k}({\bf x})$ in this region is also a decaying wave:
\begin{eqnarray}\label{eq9}
	\psi_{\rm III}({\bf x})=
	\frac{t' e^{ik_x x}}{\sqrt{4 L_x L_y}}
	\begin{pmatrix} 
		D_{(k l_B)^2/2-1}[-\sqrt{2}((y+L_y)/l_B-k_x l_B)]\\
		\frac{\sqrt{2}}{k l_B} D_{(k l_B)^2/2}[-\sqrt{2}((y+L_y)/l_B-k_x l_B)] 
	\end{pmatrix}	.
\end{eqnarray}
The eigenstate $\psi_{\bf k}({\bf x})$ is normalized due to $\int_{-\infty}^{+\infty} \int_{0}^{L_x} \psi^{\dagger}_{\bf k}({\bf x}) \psi_{\bf k}({\bf x}) \mathrm{d} x \mathrm{d} y=1$ in the limit $L_y \rightarrow +\infty$. 

Now, we are going to determine the value of $E_{\bf k}$ by enforcing the continuity of the eigenstate at $y=0$ and $y=-L_y$.
For $\psi_{\rm I}(x,0)=\psi_{\rm II}(x,0)$ at $y=0$, we get  
\begin{eqnarray}\label{eq10}
	t =
	-\frac{2 i \sin \theta e^{i\theta/2}}{D_1+D_2e^{i\theta}},
\end{eqnarray}
\begin{eqnarray}\label{eq11}
	r(k,\theta)
	\equiv e^{i\phi_r(k,\theta)} 
	=
	-\frac{D_2+D_1e^{i\theta}}{D_1 +D_2e^{i\theta}}
	,
\end{eqnarray}
where $D_1=D_{(k l_B)^2/2-1}(-\sqrt{2} k_x l_B)$ and $D_2=\sqrt{2}/(k l_B) D_{(k l_B)^2/2}(-\sqrt{2} k_x l_B)$. $\phi_r(k,\theta)$ is the reflection phase, representing the phase shift between the reflected wave and the incident wave. At $y=-L_y$, $\psi_{\rm I}(x,-L_y)=\psi_{\rm III}(x,-L_y)$, so we get
\begin{eqnarray}\label{eq12}
	r(k,\theta)e^{i 2 k_y L_y}
	\equiv e^{i\phi'_r(k,\theta)} 
	=
	-\frac{D'_2+D'_1e^{i\theta}}{D'_1 +D'_2e^{i\theta}}
	,
\end{eqnarray}
where $D'_1=D_{(k l_B)^2/2-1}(\sqrt{2} k_x l_B)$ and $D'_2=-\sqrt{2}/(k l_B) D_{(k l_B)^2/2}(\sqrt{2} k_x l_B)$. Combining the Eq.~\ref{eq11} and the Eq.~\ref{eq12}, we get
\begin{eqnarray}\label{eq13}
	k_y
	=
	\frac{n\pi}{L_y}+\frac{\phi'_r(k,\theta)-\phi_r(k,\theta)}{2L_y} \quad (n \in \mathbb{Z})
\end{eqnarray}
For a given $k_x$, both $\theta$ and $k$ is the function of $k_y$. Therefore, one will get the values of $k_y$ by solving the Eq.~\ref{eq13}. This equation can be analytically solved in the limit $L_y \rightarrow +\infty$. We define $k_n$ as the root of Eq.~\ref{eq13} for a given $n$, and $\delta_n=k_n-k_{n-1}$. Then,
\begin{eqnarray}\label{eq14}
	\delta_n
	&=&
	\frac{\pi}{L_y}+\frac{\phi'_r(k(k_n),\theta(k_n))-\phi_r(k(k_n),\theta(k_n))}{2L_y} -\frac{\phi'_r(k(k_{n-1}),\theta(k_{n-1}))-\phi_r(k(k_{n-1}),\theta(k_{n-1}))}{2L_y} \nonumber \\
	&=&\frac{\pi}{L_y}+ \left ( \frac{\mathrm{d} \phi'_r(k(k_n),\theta(k_n)) }{\mathrm{d} k_n}-\frac{\mathrm{d} \phi_r(k(k_n),\theta(k_n)) }{\mathrm{d} k_n} \right )\frac{\pi}{2L_y^2}
	+ O(1/L_y^2)
	.
\end{eqnarray}
In the limit $L_y \rightarrow +\infty$, $\delta_n=\pi/L_y$ and so $k_n=n\pi/L_y+k_0$. Obviously, $k_0=0$ is one root of the Eq.~\ref{eq13}. So $k_n=n\pi/L_y$. Note that $\psi_{k_x,-k_y}({\bf x})=r(k,\theta)^{-1}\psi_{k_x,k_y}({\bf x})$ which means $\psi_{k_x,k_y}({\bf x})$ and $\psi_{k_x,-k_y}({\bf x})$ represent the same eigenstate, so $k_y$ only takes the positive values with
\begin{eqnarray}\label{eq15}
	k_y
	=
	\frac{n\pi}{L_y} \quad (n \in \mathbb{N}).
\end{eqnarray}
In summary, the eigenstate is:
\begin{eqnarray}\label{eq151}
	\psi_{\bf k}({\bf x})=  
	\begin{cases}
		\psi_{\rm II}({\bf x}),&y \geq 0 \\
		\psi_{\rm I}({\bf x}),&-L_y< y<0 \\
		\psi_{\rm III}({\bf x}),&y\leq-L_y \\
	\end{cases}	.
\end{eqnarray}
and the eigenvalue $E_{\bf k}=\pm\hbar v_F \sqrt{k_x^2+k_y^2}$.  $(k_x,k_y)=(2\pi m/L_x,\pi n/L_y)$ with $m\in \mathbb{Z}$ and $n\in \mathbb{N}$ as $L_y \rightarrow +\infty$. Note that the $\psi_{\rm III}({\bf x})$ has no contribution for the calculation of equilibrium current and the number of chiral channels, so we do not show the expression of $\psi_{\rm III}({\bf x})$ in the main text.

\section{Half-integer quantized chiral channels}
\begin{figure}[htbp]
	\includegraphics[scale=0.4]{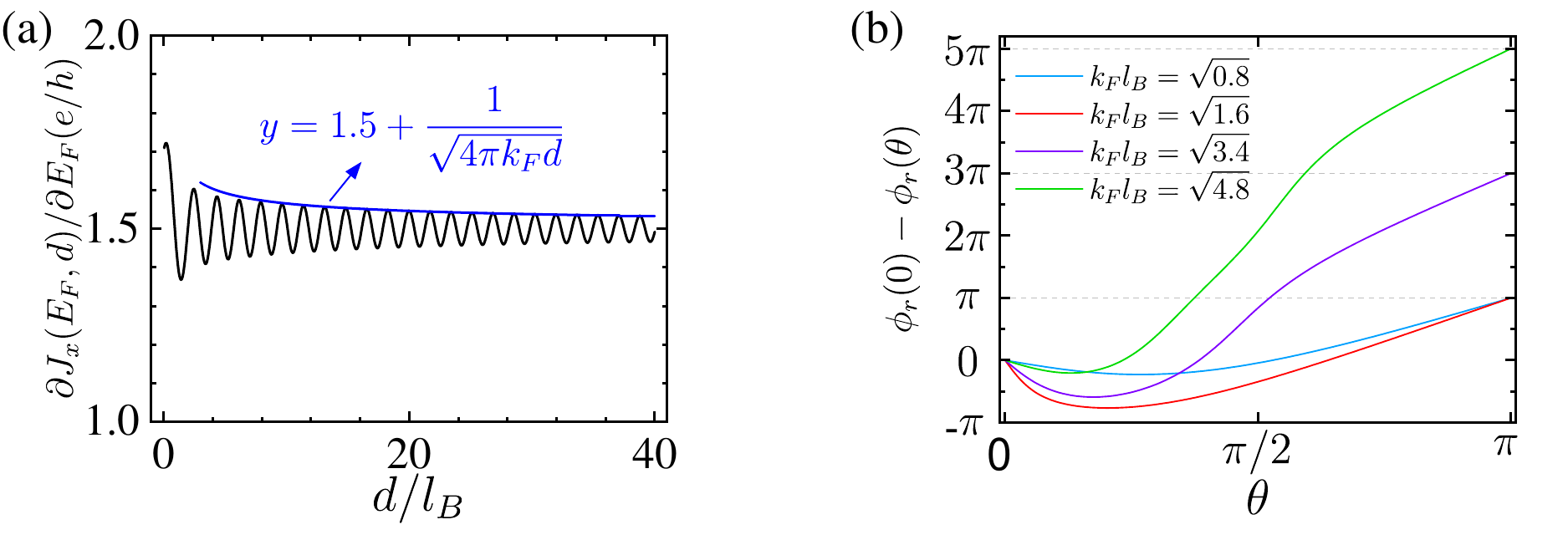}
	\caption{(color online) 
		(a) ${\partial J_y(E_F,d)}/{\partial E_F}$ as a function of the distance $d$ with $k_F l_B=\sqrt{3.4}$. 
		The blue line is the function of $y=1.5+1/\sqrt{4\pi k_F d}$.		
		(b) The reflection phase difference as a function of the incident angle $\theta$. 
	}
	\label{figs1} 
\end{figure}

In this section, we will calculate the equilibrium current and determine the number of chiral channels (CCs). For 2D Dirac fermions, the electric current density in the $x$ direction for $\psi_{\bf k}({\bf x})$ is given by $j_{x,{\bf k}}({\bf x})=-ev_F \psi_{\bf k}^{\dagger}({\bf x}) \sigma_x \psi_{\bf k}({\bf x})$. Then, the equilibrium current flowing through the section $y>-d$ $(0<d<L_y)$ for the occupied states is $J_x(E_F,d)=\sum_{E_{\bf k}\leq E_F}
\int_{-d}^{+\infty}j_{x,{\bf k}}({\bf x})\mathrm{d}y$.
We split the integral into two parts and calculate them separately:
\begin{eqnarray}\label{eq16}
	\int_{-d}^{+\infty}j_{x,{\bf k}}({\bf x})\mathrm{d}y
	&=
	-\int_{-d}^{0}
	ev_F \psi_{{\rm I},\bf k}^{\dagger}({\bf x}) \sigma_x \psi_{{\rm I},\bf k}({\bf x})
	\mathrm{d}y
	-
	\int_{0}^{+ \infty}
	ev_F \psi_{{\rm II},\bf k}^{\dagger}({\bf x}) \sigma_x \psi_{{\rm II},\bf k}({\bf x})
	\mathrm{d}y
	.
\end{eqnarray}
For the first part, we use Eq.~\ref{eq5} and obtain
\begin{eqnarray}\label{eq17}
	\int_{-d}^{0}
	ev_F \psi_{{\rm I},\bf k}^{\dagger}({\bf x}) \sigma_y \psi_{{\rm I},\bf k}({\bf x})
	\mathrm{d}y
	&=\frac{e v_F}{2 L_x L_y} \left [ 2 d \cos \theta +\frac{\sin (\phi_r +2dk_y)-\sin\phi_r}{k_y} \right ]
	.
\end{eqnarray}
For the second part, we use Eq.~\ref{eq7} and obtain
\begin{eqnarray}\label{eq18}
	\int_{0}^{+ \infty}
	ev_F \psi_{{\rm II},\bf k}^{\dagger}({\bf x}) \sigma_y \psi_{{\rm II},\bf k}({\bf x})
	\mathrm{d}y
	&=&-\frac{e v_F |t|^2}{2 L_x L_y} \frac{\sqrt{2}}{kl_B}
		\int_{0}^{+ \infty}	
		D_{(k l_B)^2/2-1}[\sqrt{2}(y/l_B-k_x l_B)]
		D_{(k l_B)^2/2}[\sqrt{2}(y/l_B-k_x l_B)]
		\mathrm{d}x \nonumber \\
		&=&-\frac{e v_F |t|^2 l_B^2}{4 L_x L_y}
		\left[ k (D_1^2+D_2^2)+2 k_x  D_1 D_2 \right] \nonumber
		\\
		&=&\frac{e v_F}{2 L_x L_y}
		\left [ \frac{\partial \phi_r(k,\theta)}{k \partial \theta}
		+\frac{\sin \phi_r}{k_y} \right ]
\end{eqnarray}
where we use the relation
\begin{eqnarray}\label{eq19}
	\frac{\partial \phi_r(\theta)}{k \partial \theta}
		&=&
		\frac{D_1^2-D_2^2}{k |D_1+D_2e^{i\theta}|^2}-\frac{2\sin^2 \theta l_B^2 }{|D_1+D_2 e^{i\theta}|^2}[k (D_1^2+D_2^2)+2 k_x D_1 D_2],
\end{eqnarray}
and
\begin{eqnarray}\label{eq20}
	\frac{\sin \phi_r}{k_y}
	&=&
	-\frac{D_1^2-D_2^2}{k|D_1+D_2e^{i\theta}|^2}
	.
\end{eqnarray}
Then, we calculate $J_x(E_F,d)$ by converting the sum into an integral as $L_x$ and $L_y$ approach to infinity:
\begin{eqnarray}\label{eq21}
	J_x(E_F,d)
	&=&-
	\lim_{L_x  \rightarrow +\infty} 
	\lim_{L_y \rightarrow +\infty}
	\sum_{E_{\bf k}\leq E_F}
    \frac{e v_F}{2 L_x L_y}
    \left [ 2 d \cos \theta  +\frac{\partial \phi_r(k,\theta)}{k \partial \theta}+
    \frac{\sin (\phi_r +2dk_y)}{k_y}
    \right ] \nonumber \\
    &=&
    -\int_{0}^{+\infty}
    \frac{\mathrm{d} k_y}{\pi/L_y} 
    \int_{-\infty}^{+\infty} 
    \frac{\mathrm{d} k_x}{2\pi/L_x}
    \Theta(E_F-E_{\bf k})
    \frac{e v_F}{2 L_x L_y}
    \left [ 2 d \cos \theta  +\frac{\partial \phi_r(k,\theta)}{k \partial \theta}+
    \frac{\sin (\phi_r +2dk_y)}{k_y}
    \right ]\nonumber \\
    &=&
    -\int_{0}^{+\infty}
    \mathrm{d} k
    \int_{0}^{\pi} 
    k \mathrm{d} \theta
    \Theta(E_F-E_{\bf k})
    \frac{e v_F}{(2 \pi)^2}
    \left [ 2 d \cos \theta  +\frac{\partial \phi_r(k,\theta)}{k \partial \theta}+
    \frac{\sin (\phi_r +2dk_y)}{k_y}
    \right ]
\end{eqnarray}
where the $\Theta(x)$ is the step function. 
$J_x(E_F,d)$ is nonzero and exhibits a chiral nature near the edge of the metallic region due to the breaking of time-reversal symmetry in the presence of a magnetic field.
By taking the partial derivative of $J_x(E_F,d)$, one can get
\begin{eqnarray}\label{eq22}
	\frac{\partial J_x(E_F,d)}{\partial E_F}
	&=&
	-\int_{0}^{+\infty}
	\mathrm{d} k
	\int_{0}^{\pi} 
	k \mathrm{d} \theta
	\delta(E_F-E_{\bf k})
	\frac{e v_F}{(2 \pi)^2}
	\left [ 2 d \cos \theta  +\frac{\partial \phi_r(k,\theta)}{k \partial \theta}+
	\frac{\sin (\phi_r +2dk_y)}{k_y}
	\right ] \nonumber \\
	&=&
	\frac{e}{h} 
	\int_{0}^{\pi} 
	\frac{\mathrm{d} \theta}{2\pi}
	\left ( -
	\frac{\partial \phi_r(k_F ,\theta)}{ \partial \theta}
	\right )
	-
	\frac{e}{h} 
	\int_{0}^{\pi} 
	\frac{\mathrm{d} \theta}{2\pi}
	\frac{\sin (\phi_r +2dk_F \sin \theta)}{\sin \theta}
	.
\end{eqnarray}
Here, we use the relation $E_{\bf k}=\hbar v_F k$ and  define $ k_F =E_F/(\hbar v_F)$.
$[(h/e){\partial J_x(E_F,d)}/{\partial E_F}]$ is the number of CCs in the region $y>-d$, which directly determines the Hall transport. 
The first term of Eq.\ref{eq22} is independent of the distance $d$.
 The second term of the Eq.~\ref{eq22} is approximately equal to $1/\sqrt{4\pi k_F d} \cos (2k_F d +\phi_0(k_F))$ as $ d/l_B $ tends to the infinity [see Fig.~\ref{figs1}(a)]. 
 So this term decays to zero in a power law $(d/l_B)^{-1/2}$, which means that the equilibrium current is mainly located at the boundary between the metallic region ($y<0$) and the magnetic field region ($y>0$). 
 
 For convenience, we define $J_c(E_F)=\lim_{d  \rightarrow +\infty} J_x(E_F,d)$ as the $d$-independent part of the equilibrium current $J_x(E_F,d)$. $J_c(E_F)$ is the total equilibrium current near the edge of metallic region.
 Therefore, $[(h/e){\partial J_c(E_F)}/{\partial E_F}]$ is the total number of CCs located at edge of the metallic region. Then, we obtain
\begin{eqnarray}\label{eq23}
  	\frac{h}{e}\frac{\partial J_c(E_F)}{\partial E_F}
  	&=&
  	\int_{0}^{\pi} 
  	\frac{\mathrm{d} \theta}{2\pi}
  	\left ( -
  	\frac{\partial \phi_r(k_F ,\theta)}{ \partial \theta}
  	\right ) \nonumber \\
  	&=&
  	\frac{\phi_r(k_F ,0)-\phi_r(k_F , \pi)}{2\pi}  	
  	.
  \end{eqnarray}
The relationship $r(k_F,0)r(k_F,\pi)=-1$ derived from Eq.~\ref{eq12} leads to $\phi_r(k_F ,0)-\phi_r(k_F , \pi)=2\pi(n+1/2)$ with $n\in \mathbb{Z}$.
The exact value of $n$ is determined by the Fermi energy $E_F$ [see Fig.~\ref{figs1}(b)], and $n=\lfloor (k_F l_B)^2/2 \rfloor=\lfloor ( {E_F}/{\epsilon_D})^2/2 \rfloor$ where $\lfloor x \rfloor$ denotes the largest integer less than $x$. So
\begin{eqnarray}\label{eq24}
		\frac{h}{e}\frac{\partial J_c(E_F)}{\partial E_F}	
	&=&
	 \lfloor ( {E_F}/{\epsilon_D})^2/2 \rfloor+\frac{1}{2}.  	
\end{eqnarray}
Half-integer quantized $\partial J_c(E_F)/\partial E_F$ implies that there are half-integer quantized CCs near the edge of the magnetic field region, which is the origin of the half-integer Hall conductance.

\section{The relation between the CCs and the Hall conductance  }
\begin{figure}[htbp]
    \includegraphics[scale=0.4]{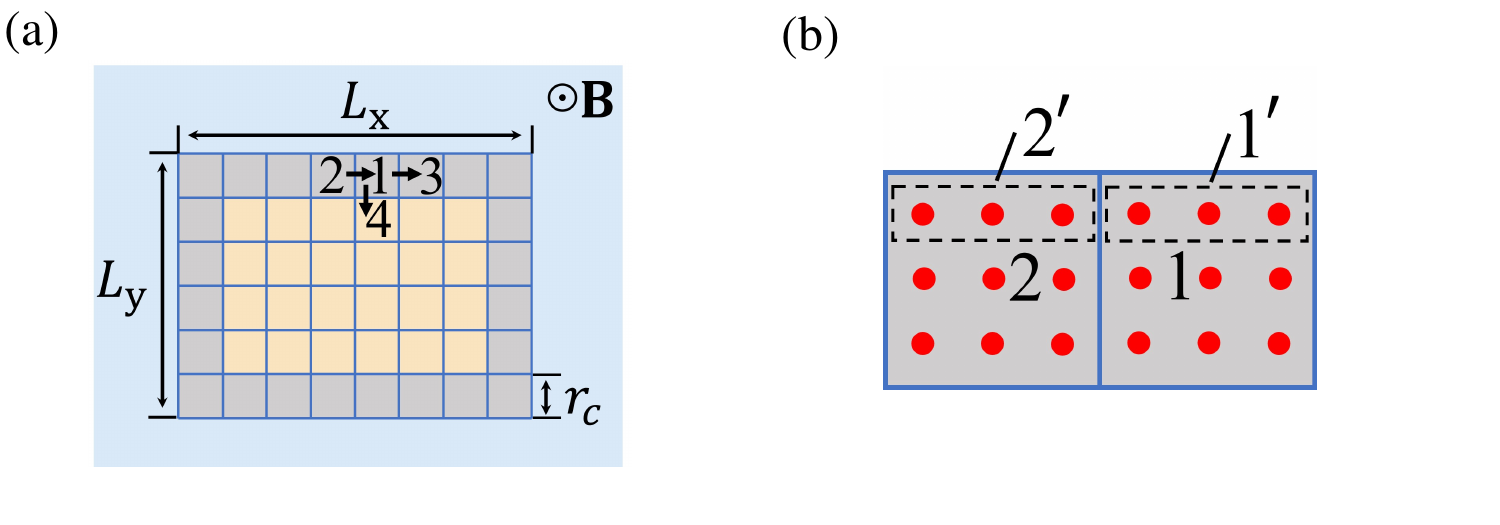}
    \caption{(color online) (a) Schematic plot of the 2D Dirac fermions with a perpendicular magnetic field applied in the blue region. The yellow and gray region are metallic and divided into many boxes with size of $r_c$. The number are used to label boxes, and the black arrow represents the direction of the net current. 
    (b) The red balls represent virtual leads, and $1'$ and $2'$ represent the dashed black boxes.
    }
    \label{figs2}
  \end{figure}
In this section, we will establish an analytic relation between the CCs and the Hall conductance. 
We consider a metallic system (yellow region) with half-integer quantized CCs circulating around its edge, as showed in Fig.~\ref{figs2}(a).
To observe a stable HIHC in experiments,  it is essential for both the Hall resistivity and longitudinal resistivity to remains independent of system's size.
For this purpose, we assume an Ohmic scaling of the conductance $G_{\text{2D}}$ in the metallic system, i.e. $G_{\text{2D}}=\sigma_nL_y/L_x$, ensuring that the conductivity $\sigma_n$ is unaffected by the system's size.
In this scenario, we can consider this metallic system as composed of numerous independent boxes, each with size $r_c$ and conductance $G=\sigma_n$. These boxes can be viewed as independent classical resistors connected together in accordance with Ohm's law.
To incorporate the contribution from the CCs near the edge, we introduce the 1D conductivity $\sigma_{\text{1D}}$ that arises from the CCs, and assume the corresponding conductance satisfies 1D Ohmic scaling law, i.e. $G_{\text{1D}}=\sigma_{\text{1D}}/L_x$.
We assume that there is a slowly varying electric potential with a uniform electric field ${\bf E}=(E_x,-E_y)$ in metallic region.
The electric voltage distribution is $V=-x E_x + y E_y$ for given ${\bf E}=(E_x,-E_y)$.

Considering the current conservation in each gray box such as the box 1, we obtain:
\begin{eqnarray}\label{eq25}
	[\sigma_n E_x r_c+\sigma_{\text{1D}} E_x+J_c(eV_{12})]=
	[\sigma_n E_x r_c+\sigma_{\text{1D}} E_x+J_c(eV_{13})]+E_y \sigma_n r_c
\end{eqnarray}
The left-hand side represents the coming current, while the right-hand side represents the outgoing current. The  left-hand side is the current that flows from box 2 to box 1. $[\sigma_n E_x r_c+\sigma_{\text{1D}} E_x]$ is the longitudianal current, and $J_c(eV_{12})$ is the Hall current flowing through the CCs. Here, $V_{12}$ is the voltage at the boundary between the box 1 and the box 2, and similarly for $V_{13}$. The first term of the right-hand side is the longitudinal current flowing from the box 1 to the box 3. The second term of the left-hand side is the current flowing from the box 1 to the box 4 thought the normal channels. 
Considering the voltage varies very slowly with space ($|eV-E_F|\ll \epsilon_D$), we get $J_c(eV)=\partial J_c(E_F) / \partial E_F (eV-E_F) +J_c(E_F)$. So the Eq.~\ref{eq25} becomes:
\begin{eqnarray}\label{eq26}
 E_y \sigma_n=E_x e \frac{\partial J_c(E_F)}{\partial E_F}
\end{eqnarray}
Additionally, the total current flowing in the horizontal ($x$) direction is:
\begin{eqnarray}\label{eq27}
 I_x &=& \sigma_n E_x L_y+\sigma_{\text{1D}} E_x l_{\phi}+J_c(eV_{\text{up}})-J_c(eV_{\text{down}}) \nonumber \\
 &=& \sigma_n E_x L_y+\sigma_{\text{1D}} E_x+ e \frac{\partial J_c(E_F)}{\partial E_F} E_y L_y
\end{eqnarray}
where the $V_{\text{up}}$($V_{\text{down}}$) is the voltage of the upper (lower) edge,  and $V_{\text{up}}-V_{\text{down}}=E_y L_y$. 
the first two terms are the longitudinal current, and the third term is the Hall current from the CCs.
Then the longitudianal resistance $\rho_{xx}=E_x L_y/I_x$ and Hall resistance $\rho_{xy}=E_y L_y/I_x$. 
The longitudinal conductance $\sigma_{xx}$ and Hall conductance $\sigma_{xy}$ are obtained from the tensor relation: $\sigma_{xx}=\rho_{xx}/(\rho_{xx}^2+\rho_{xy}^2)$, and $\sigma_{xy}=\rho_{xy}/(\rho_{xx}^2+\rho_{xy}^2)$.
\begin{eqnarray}\label{eq28}
	\sigma_{xy}(E_F)=\left[1+\frac{1}{1+(\sigma_c/\sigma_n)^2}\frac{\sigma_{\text{1D}}}{ \sigma_n L_y }\right] \sigma_c \nonumber \\
		\sigma_{xx}(E_F)= \left[1+\frac{1}{1+(\sigma_c/\sigma_n)^2}\frac{\sigma_{\text{1D}}}{\sigma_n L_y}\right]\sigma_n
\end{eqnarray}
where we define $\sigma_c \equiv e\partial J_c(E_F) / \partial E_F$.

\section{Dissipation from the CCs}
In this section, we will demonstrate how the dissipation emerges in the CCs. we consider $r_c^2$ virtual leads attached on the box 1 and box 2 in Fig.~\ref{figs2}(b). We assume that the transition coefficient between the lead $p$ and the $q$ is $T_{pq}$ and the CCs are located at the edge. 
The voltage of each virtual leads is $V({\bf r})=-E_x x +E_y y$,  where ${\bf r}_p=(x_p,y_p)$ is the coordinate of the virtual lead. 
We denotes $V_{\text{up}}=-E_x x_0+E_y y_0$ where $(x_0,y_0)$ is the space coordinate of the virtual lead in the top left corner of the box 1.
Then the current flowing from the box 1 to the box 2 is
\begin{eqnarray}\label{eq281}
	I_{21}&=&\frac{e^2}{h}\sum_{p\in 2,q\in 1}\left(T_{pq} V_q-T_{qp}V_p\right) \nonumber \\
	&=&\frac{e^2}{h}\sum_{p\in 2,q\in 1}\left(T_{pq} -T_{qp}\right)V_q+\frac{e^2}{h}\sum_{p\in 2,q\in 1}T_{qp}(V_q-V_p) \nonumber \\
	&=&\frac{e^2}{h}\sum_{p\in 2',q\in 1'}\left(T_{pq} -T_{qp}\right)(x_0-x_q)E_x+\frac{e^2}{h}\sum_{p\in 2',q\in 1'}\left(T_{pq} -T_{qp}\right)V_{\text{up}}+\frac{e^2}{h}\sum_{p\in 2,q\in 1}T_{qp}(x_p-x_q)E_x \nonumber \\
	&=&\sigma_{\text{1D}}E_x+\sigma_c V_{\text{up}}+\sigma_n E_x
\end{eqnarray}
Here, $\sigma_{\text{1D}}={e^2}/{h}\sum_{p\in 2',q\in 1'}\left(T_{pq} -T_{qp}\right)(x_0-x_q)$ obviously comes from the chiral channels located the edge, which provides the longitudinal current. $\sigma_c={e^2}/{h}\sum_{p\in 2',q\in 1'}\left(T_{pq} -T_{qp}\right)$ represents the half-integer quantized chiral channels, which provides the Hall currents. Meanwhile, $\sigma_n={e^2}/{h}\sum_{p\in 2,q\in 1}T_{qp}(x_p-x_q)$ represents the normal channels, which provides the longitudinal currents.
There is a voltage drop along the CCs when the CCs coexist with the metallic system, which leads to dissipation in the CCs. So we refer to the CCs as dissipative CCs.

\section{Lattice model Hamiltonian and dephasing}
\begin{figure}[htbp]
	\includegraphics[scale=0.4]{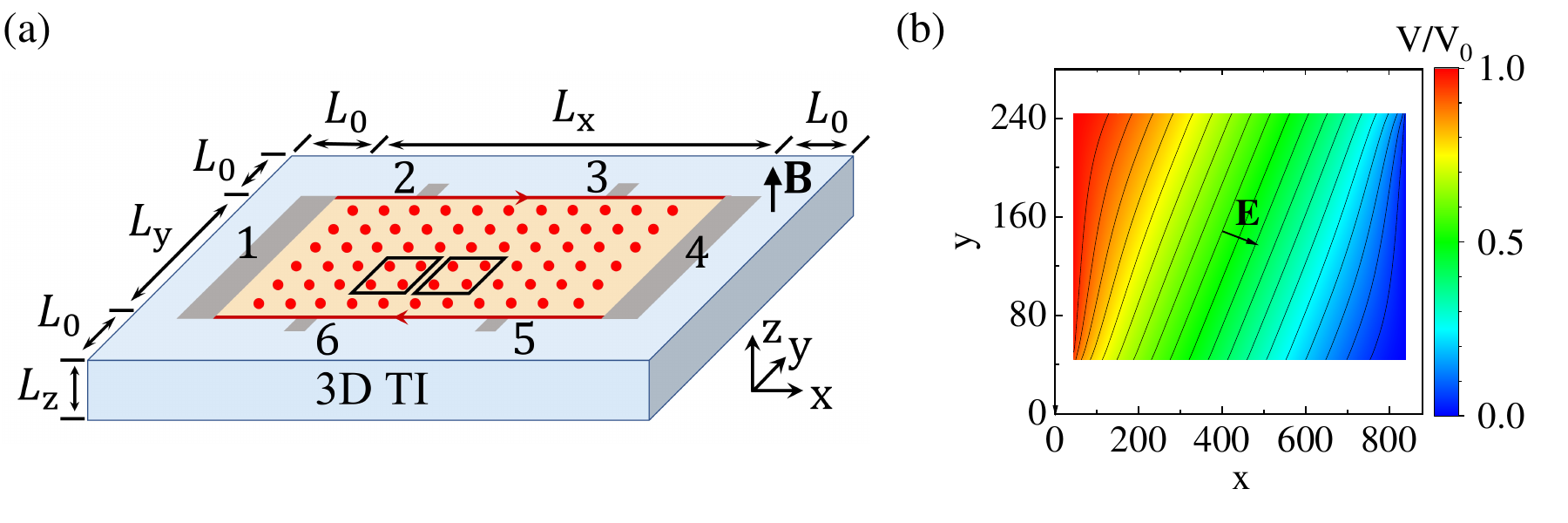}
	\caption{(color online) 
		(a) Schematic plot of the 3D TI with a perpendicular magnetic field applied in the blue region. The yellow region on the top surface is a Dirac metal, and the red balls represent the sites attached to virtual leads. The size of the two adjacent black boxes is $r_x \times r_y$. The numbers are used to label the real leads.
		(b) The space distribution of the voltage $V$ with $\Gamma_v=4.0$ and $E_F=0.42$.	$L_x \times L_y \times L_z=800 \times 201 \times 6$, $L_0=40$, $n_x \times n_y=200 \times 51$.	Other parameters are the same as those in Fig.~2 in the main text.
	}
	\label{figs3}
\end{figure}

We consider a 3D topological insulator (TI) with a perpendicular magnetic field applied in the blue region as showed in the Fig.~\ref{figs3}(a). 
Here, ${\bf B}=B(x,y)\hat{e}_z$, and $B(x,y)=0$ in the yellow region ($L_0<x<L_0+L_x$, and $L_0<y<L_0+L_y$), and $B(x,y)=B_0$ elsewhere. We choose the vector potential in the gauge ${\bf A}=A(x,y)\hat{e}_x$. $A(x,y)=-yB_0$ for $0<x<L_0$ and $L_0+L_x<x<2L_0+L_x$; For $L_0<x<L_0+L_x$, we have: 
\begin{eqnarray}\label{eq30}
	A(x,y)=  
	\begin{cases}
		-yB_0,&0<y<L_0 \\
		-L_0 B_0,&L_0\leq y\leq L_0+L_y \\
		-(y-L_y)B_0,&L_0+L_y <y< 2L_0+L_y\\
	\end{cases}	.
\end{eqnarray}

We discretize the Hamiltonian $H_{\text{3D}}$ in the main text on cubic lattices:
\begin{eqnarray}\label{eq31}
	H_{\text{3D}}=&\left[
	\sum_{{\bf x}}{\bf c}_{\bf x}^{\dagger}{\mathcal M}_0{\bf c}_{\bf x}
	+\left(\sum_{{\bf x},j=x,y,z}{\bf c}_{\bf x}^{\dagger}{\mathcal T}_{{\bf x},j}{\bf c}_{{\bf x}+{\bf \hat{e}}_j}+\text{H.c.}\right)
	\right]
 +H_{\text{lead}}
\end{eqnarray}
where ${\mathcal M}_0=\left(M_0-6B/a^2\right)\sigma_z{\otimes}s_0$, and ${\mathcal T}_{{\bf x},j}=(B/a^2\sigma_z{\otimes}s_0-iA/(2a)\sigma_x{\otimes}s_j)e^{i(e/\hbar)a A(x,y)\delta_{y,j}}$.
The magnetic field ${\bf B}$ is included by doing the Peierls
substitution \cite{Hofstadter1976} as a phase of the hopping term.
Here, $a=1$ is the cubic lattice constant.
${\bf c}_{\bf x}$ (${\bf c}_{\bf x}^{\dagger}$) is the annihilation (creation) operator at site ${\bf x}$, and ${\bf \hat{e}}_j$ is a unit vector in the $j$ direction for $j=x,y,z$.
$\delta_{y,j}=1$ for $j=y$ and is zero for $j=x,z$.
The first and the second terms in Eq.~\ref{eq31} describe the 3D TI in the central region. The last term $H_{\text{lead}}=
\sum_{{\bf x},{\bf k}}\left[\epsilon_{{\bf k}}{\bf a}_{{\bf x},{\bf k}}^{\dagger}{\bf a}_{{\bf x},{\bf k}}+(t_{{\bf x},{\bf k}}{\bf a}_{{\bf x},{\bf k}}^{\dagger}{\bf c}_{{\bf x}}+\text{H.c.})\right]$ represents the Hamiltonian of virtual leads and real leads, and their couplings to the central sites.

Then, we can calculate the transmission coefficient $T_{pq}(E_F)$.
$T_{pq}(E_F)=\mbox{Tr}[{\bm \Gamma}_p{\bm G}^r{\bm \Gamma}_q{\bm G}^a]$, where the linewidth function ${\bm \Gamma}_p=i\left({\bm \Sigma}_p^r-{\bm \Sigma}_p^{r\dagger}\right)$ and the Green's function ${\bm G}^r=[{\bm G}^a]^{\dagger}=[E_F{\bm I}-H_{\mbox{cen}}-\sum_{p}{\bm \Sigma}_p^{r}]^{-1}$. ${\bm \Sigma}_p^r$ is the retarded self-energy due to the coupling to the lead $p$ and $H_{\mbox{cen}}$ is the lattice Hamiltonian of the TI [the first and the second term of Eq.~\ref{eq31}].
Note that we use the wide-band approximation for the real leads in the main text with ${\bm \Gamma}_p=\Gamma_v {\bf I}_p$, and $\Gamma_v=2\pi \rho t_k^2$  is the dephasing strength and $\rho$ is the density of state in the virtual lead \cite{Y2008}.
Only electrons in the gapless surface states participate in the electrical transport when $E_F$ is located within the Landau levels at zero or low temperature. Therefore, we just need to consider the dephasing process on the metallic region.
We simulate the dephasing process by using $n_x\times n_y$ uniformly distributed B{\"u}ttiker's virtual leads on the metallic region [yellow region in Fig.~\ref{figs3}(a)].

In our calculations, the real lead 1 and lead 4 act as current electrodes and other real leads act as voltage electrodes. Thus, the longitudinal current $I_1=-I_4\equiv I_x$ when a external bias is applied between lead 1 and lead 4 with $V_1=V_0$ and $V_4=0$. Virtual leads act as dephasing, so the net current of each virtual lead is zero. 
Combining Landauer-B{\"u}ttiker formula\cite{Buttiker1988,Data1995} with those boundary conditions in the real and virtual leads, one can calculate the voltage of each real and virtual lead and the longitudinal current $I_x$. We define $V(x,y)$ as the voltage of the virtual leads, and $(x,y)$ is the coordinate of the virtual leads. We show $V(x,y)$ varies uniformly with respect to spatial position in the central region of Fig.~\ref{figs3}(b). This indicate that the electric field ${\bf E}=-{\bm \nabla} V$ is uniform and thus Ohmic scaling of bulk conductance holds.
The longitudinal and Hall resistance are given by $\rho_{xx}=[(V_2-V_3)/I_x]/(L/W)$ and $\rho_{xy}=(V_2-V_6)/I_x$, respectively. We then obtain the longitudinal and Hall conductance by using $\sigma_{xx}=\rho_{xx}/(\rho_{xx}^2+\rho_{xy}^2)$ and $\sigma_{xy}=\rho_{xy}/(\rho_{xx}^2+\rho_{xy}^2)$. 
$L$ is the length between the lead 2 and 3, and $W=L_y$ is the width of the metallic region.

\section{The Hamiltonian of TI with the gate voltages and the Hall bar device}
In the main text, we apply two gate voltages on the front and back surface of the 3D TI, respectively[see Fig.~4(a) in the main text]. Here, we show the Hamiltonian of 3D TI when applying the gate voltages.
We choose the vector potential in the gauge ${\bf A}=A(x)\hat{e}_z=-x B_0\hat{e}_z$ and discretize the Hamiltonoan $H_{\text{3D}}$ on cubic lattices:
\begin{eqnarray}\label{eq32}
	H_{\text{3D}}=&
	\sum_{{\bf x}}{\bf c}_{\bf x}^{\dagger}{\mathcal M}_0{\bf c}_{\bf x}
	+\left(\sum_{{\bf x},j=x,y,z}{\bf c}_{\bf x}^{\dagger}{\mathcal T}_{{\bf x},j}{\bf c}_{{\bf x}+{\bf \hat{e}}_j}+\text{H.c.}\right)
	+ H_{g1}+H_{g2}
\end{eqnarray}
where ${\mathcal M}_0=\left(M_0-6B/a^2\right)\sigma_z{\otimes}s_0$, and ${\mathcal T}_{{\bf x},j}=(B/a^2\sigma_z{\otimes}s_0-iA/(2a)\sigma_x{\otimes}s_j)e^{i(e/\hbar)a A(x)\delta_{z,j}}$.
The magnetic field ${\bf B}$ is included by doing the Peierls substitution \cite{Hofstadter1976} as a phase of the hopping term.
$H_{g1}=\sum_{{\bf x} \in \text{front surfaces}}{\bf c}_{\bf x}^{\dagger}{\mathcal M}_{g1}{\bf c}_{\bf x}$ with ${\mathcal M}_{g1}=eV_{g1}\sigma_0{\otimes}s_0$ and $H_{g2}=\sum_{{\bf x} \in \text{back surfaces}}{\bf c}_{\bf x}^{\dagger}{\mathcal M}_{g2}{\bf c}_{\bf x}$ with ${\mathcal M}_{g2}=eV_{g2}\sigma_0{\otimes}s_0$.
Here, $a=1$ is the cubic lattice constant.
${\bf c}_{\bf x}$ (${\bf c}_{\bf x}^{\dagger}$) is the annihilation (creation) operator at site ${\bf x}=(x,y)$, and ${\bf \hat{e}}_j$ is a unit vector in the $j$ direction for $j=x,y,z$.
$\delta_{z,j}=1$ for $j=z$ and is zero for $j=x,y$.

\begin{figure}[htbp]
	\includegraphics[scale=0.4]{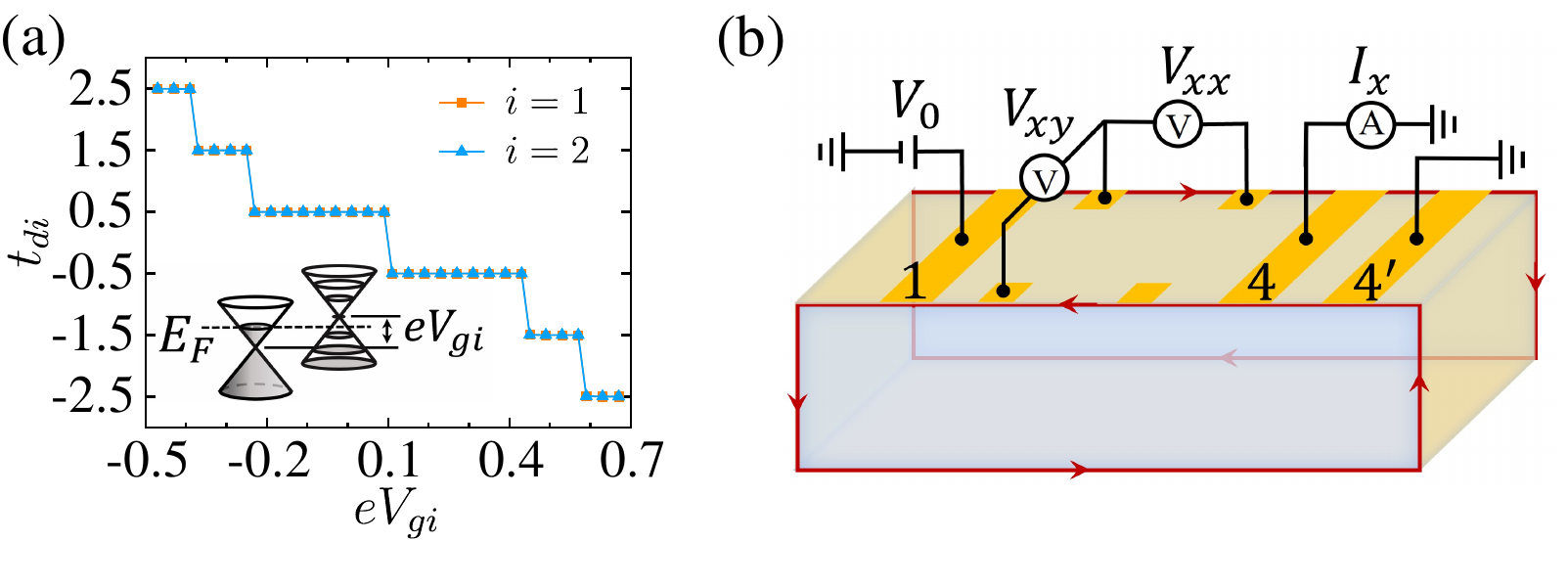}
	\caption{(color online) 
		(a) $t_{di}$ as a function of the Fermi energy $eV_{gi}$ for $i=1,2$ with $E_F=0.1$. Here, $B_0=0.01[h/(ea^2)]$, $\Gamma_v=4.0$, $N_x \times L_y \times N_z=140 \times 40 \times 40$,  $n_x \times n_y=140 \times 40$, and $r_x \times r_y =45 \times 10$.
		(b) Schematic plot of the Hall bar device. 
	}
	\label{figs4}
\end{figure}

The gate voltage $V_{g1}$ ($V_{g2}$) is applied to the front (back) surface to shift the Dirac point by an energy $eV_{g1}$ ($eV_{g2}$). Therefore, the number and direction of the CCs on the front (back) surface can be tuned by the gate voltage $V_{g1}$ ($V_{g2}$).
To demonstrate this numerically, we apply $n_x \times n_y$ virtual leads on the top surface, and define $t_{d1}\equiv T_d(n_x/2,1)$ and $t_{d2}\equiv T_d(n_x/2,n_y-r_y)$, which are the number of the CCs on the front and back surface, respectively.
Fig.~\ref{figs4}(b) shows that $t_{di}= \text{sgn}(E_F-eV_{gi})[\lfloor(E_F-eV_{gi})^2/(2\epsilon_D^2)\rfloor+1/2]$ for $i=1,2$, where $\text{sgn}(E_F-eV_{gi})$ represents the direction of the CCs.
The gate voltages $V_{g1}$ and $V_{g2}$ can introduce opposite-type (electron-hole) carriers on the respective opposite side surfaces, and lead to $t_{d1}=-t_{d2}$ when $eV_{g1}+eV_{g2}=2E_F$.
$t_{d1}=-t_{d2}$ means that there are the same number but opposite direction of CCs  located at the edge of the front and back surface, respectively.
Actually, as long as $V_{g1}$ and $V_{g2}$ satisfy the relation $\lfloor(E_F-eV_{g2})^2/(2\epsilon_D^2)\rfloor<(E_F-eV_{g2})^2/(2\epsilon_D^2)<\lfloor(E_F-eV_{g1})^2/(2\epsilon_D^2)\rfloor+1$, one can get $t_{d1}=-t_{d2}$.

When $t_{d1}=-t_{d2}$, we can obtain the HIHC by utilizing the Hall bar device to measure the longitudinal and Hall resistance of the metallic system on the top surface [see Fig.~\ref{figs4}(b)].
The current flowing out from lead 1 will be divided into two components. One component flows along the top surface and enters lead 4, while the other component flows along the bottom surface and enters lead $4'$. Since lead 4 and lead $4'$ have the same voltage, there is no net current between them. Therefore, the top and bottom surfaces can be viewed as separate entities and do not have any influence on the transport results of each other.
From this perspective, the configuration depicted in Figure~\ref{figs4}(b) is equivalent to the arrangement shown in Figure.~1(c) in the main text.

\begin{figure}[htbp]
	\includegraphics[scale=0.14]{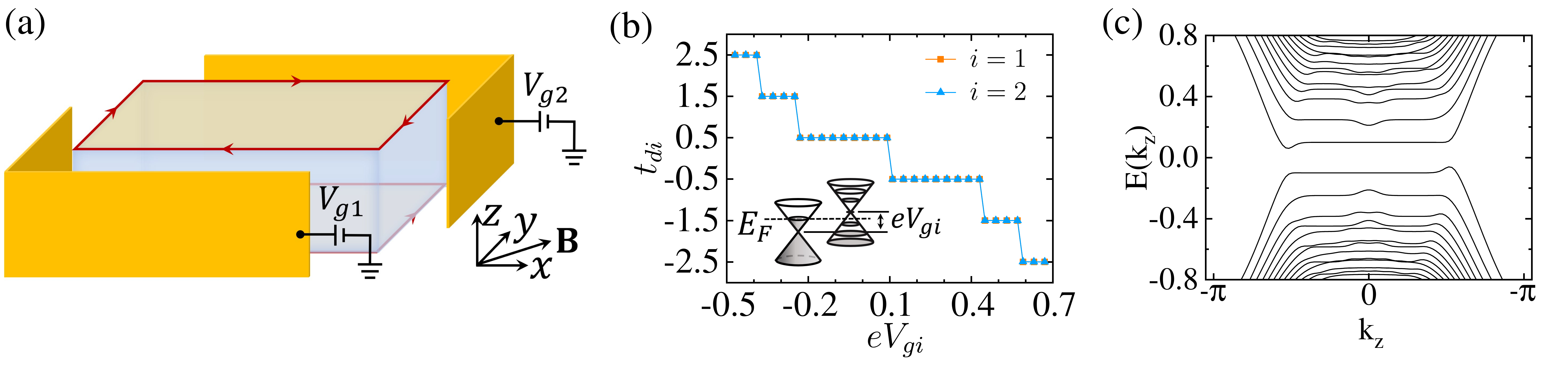}
	\caption{(color online) (a) Schematic plot of a 3D TI with a gate on the right and    back surface in the presence of a uniform magnetic field ${\bf B}=(B_0,B_0,0)$. 
		The red line represents the CCs. 
		(b) $t_{di}$ as a function of the Fermi energy $eV_{gi}$ for $i=1,2$ with $E_F=0.1$. Here, $B_0=0.01[h/(ea^2)]$, $\Gamma_v=4.0$, $L_x \times L_y \times L_z=140 \times 40 \times 40$,  $n_x \times n_y=140 \times 40$, and $r_x \times r_y =45 \times 10$.
		(c) The band structure of the 3D TI as a function of $k_z$ with $L_x \times N_y=30 \times 30$, $B_0=0.01[h/(ea^2)]$, and $eV_{g1}=-eV_{g2}=0.1$.
		Other parameters are the same as those in Fig.~(2) in the main text. 
	}
	\label{figs5}
\end{figure}

\section{An alternative approach to measure the HIHC}

In this section, we will present an alternative approach for measuring the HIHC.
In Fig.~\ref{figs5}(a), a 3D TI is subjected to a uniform magnetic field ${\bf B}=(B_0,B_0,0)$, and gate voltages are applied to the side surfaces, introducing opposite-type (electron-hole) carriers on the respective opposite side surfaces.
Due to the opposite electromagnetic responses of holes and electrons, the effective magnetic fields on the four side surfaces are directed inward towards these surfaces. 
In this scenario, the top and bottom surfaces act as Dirac metals, while being encompassed by Landau levels on the four side surface states influenced by a magnetic field. 
This  configuration  is topologically equivalent to the arrangement  shown in Fig.1(c) in the main text.
Moreover, the half-integer quantized CCs on the top and bottom surface are spatially separated by the insulating side surfaces. 
This behavior is confirmed by our numerical simulations, as shown in Fig.~\ref{figs5}(b) and (c).
Figure~\ref{figs5}(c) shows the side surfaces become gapped when the Fermi energy satisfies the relation that $td_1=-td_2$, with periodic boundary condition in the $z$ direction and the gate voltage $eV_{g1}=-eV_{g2}=0.1$.
Therefore, we can obtain the HIHC by using the six-terminal Hall bar device to measure the longitudinal and Hall resistance of the metallic system on the top surface.
Compared with the approach in the main text [see Fig.~(4) in the main text], this approach does not need the lead $4'$ in Fig.~\ref{figs4}(b). However, this approach need to apply gate on the four side surfaces, which may be more difficult to realize in experiments.

Next, we will present the details of the calculation of the band structure of the TI in a uniform magnetic field ${\bf B}=(B_0,B_0,0)$ and two different gates on the opposites side surfaces [see Fig.~\ref{figs5}(a)]. 
To make the Hamiltonian is translationally invariant along the $z$ direction, we choose the vector potential in the gauge ${\bf A}=A(x,y)\hat{e}_z=B_0(y-x)\hat{e}_z$. 
Considering the open boundary condition in the $x$ and $y$ direction and periodic boundary condition in the $z$ direction, we discretize the Hamiltonoan $H_{\text{3D}}$ on cubic lattices:
\begin{eqnarray}\label{eq32}
	H(k_z)=&
	\sum_{{\bf x}}{\bf c}_{\bf x}^{\dagger}{\mathcal M}_0{\bf c}_{\bf x}
	+\left(\sum_{{\bf x},j=x,y}{\bf c}_{\bf x}^{\dagger}{\mathcal T}_{{\bf x},j}{\bf c}_{{\bf x}+{\bf \hat{e}}_j}+\text{H.c.}\right)
	+
	\left(
	\sum_{{\bf x},j=z}
	{\bf c}_{\bf x}^{\dagger}{\mathcal T}_{{\bf x},j}{\bf c}_{{\bf x}+{\bf \hat{e}}_j}e^{ik_za}
	+
	\text{H.c.}
	\right) + H_{g1}+H_{g2}
\end{eqnarray}
where ${\mathcal M}_0=\left(M_0-6B/a^2\right)\sigma_z{\otimes}s_0$, and ${\mathcal T}_{{\bf x},j}=(B/a^2\sigma_z{\otimes}s_0-iA/(2a)\sigma_x{\otimes}s_j)e^{i(e/\hbar)a A(x,y)\delta_{z,j}}$.
The magnetic field ${\bf B}$ is included by doing the Peierls
substitution \cite{Hofstadter1976} as a phase of the hopping term.
$H_{g1}=\sum_{{\bf x} \in \text{left and front surfaces}}{\bf c}_{\bf x}^{\dagger}{\mathcal M}_{g1}{\bf c}_{\bf x}$ with ${\mathcal M}_{g1}=eV_{g1}\sigma_0{\otimes}s_0$ and $H_{g2}=\sum_{{\bf x} \in \text{right and back surfaces}}{\bf c}_{\bf x}^{\dagger}{\mathcal M}_{g2}{\bf c}_{\bf x}$ with ${\mathcal M}_{g2}=eV_{g2}\sigma_0{\otimes}s_0$.
Here, $a=1$ is the cubic lattice constant.
${\bf c}_{\bf x}$ (${\bf c}_{\bf x}^{\dagger}$) is the annihilation (creation) operator at site ${\bf x}=(x,y)$, and ${\bf \hat{e}}_j$ is a unit vector in the $j$ direction for $j=x,y,z$.
$\delta_{z,j}=1$ for $j=z$ and is zero for $j=x,y$.
By exactly diagonalizing $H(k_z)$, we obtain the band structure as showed in Fig.~\ref{figs5}(c).
This demonstrates that the side surfaces are insulating and the top and bottom surfaces are spatially separated by the insulating side surfaces when the Fermi energy satisfies the relation that $t_{d_1}=-t_{d_2}$.

\bibliographystyle{apsrev4-1}
\bibliography{suppref.bib}